\documentclass[
reprint,
superscriptaddress,
 amsmath,amssymb,
 aps,
floatfix,
nofootinbib
]{revtex4-2}

\usepackage[utf8]{inputenc}
\usepackage[english]{babel}
\usepackage{diagbox}
\usepackage{makecell}
\usepackage{textcomp}
\usepackage{amsmath,etoolbox}
\usepackage{amsthm}
\usepackage{amssymb}
\usepackage[left=2.5cm,right=2.5cm,top=3cm,bottom=3cm,bindingoffset=0cm]{geometry}
\usepackage{dsfont}
\usepackage{bm}
\usepackage{leftidx}
\usepackage{float}
\usepackage[makeroom]{cancel}
\usepackage{indentfirst}
\usepackage{underoverlap}
\usepackage{caption}
\usepackage{hyperref}
\usepackage{cleveref}
\usepackage[table]{xcolor}
\usepackage{tikz}
\usetikzlibrary{quantikz,decorations.pathmorphing, patterns}
\usepackage{subfig}
\usepackage{graphicx}
\usepackage[titletoc,title]{appendix}

\usepackage{adjustbox}
\usepackage{mathtools}
\usepackage{multirow}
\usepackage{xargs}
\usepackage{xstring}
\usepackage{rotating}
\usepackage{calc}
\usepackage{slashbox}
\usepackage{upgreek}
\usepackage{enumitem}
\usepackage{pst-node}
\usepackage{tikz-cd}

\usetikzlibrary{matrix,arrows,decorations.pathmorphing}
\newenvironment{sloppypar*}{\sloppy\ignorespaces}{\par}
\allowdisplaybreaks %

\newcommand {\sket} [1] {| #1 \rangle}
\newcommand {\bket} [1] {\bigl| #1 \bigr\rangle}
\newcommand {\sbraket} [2] {\langle #1 | #2 \rangle}

\newcommand {\sand} [3] {\langle #1 | #2 | #3 \rangle}

\newcommand\abs[1]{\left|#1\right|}
\newcommand\spec{\operatorname{spec}}

\newcommand {\unit} {\mathds{1}}

\DeclareMathOperator{\diag}{diag\s{}  }

\DeclareMathOperator{\Span}{span}
\DeclareMathOperator{\rank}{rank}
\newcommand{\bma} {\begin{pmatrix}}
\newcommand{\ema} {\end{pmatrix}}

\newcommand{\iu}{{i\mkern1mu}}

\renewcommand{\d}[1]{\ensuremath{\operatorname{d}\!{#1}}}
\newcommand{\me}{\mathrm{e}}
\newcommand{\s}{\hspace{.08em}}

\newcommandx*\hilbdimK[1][1=]{\mathcal{D}_{K\IfEqCase{#1}{{}{}}[=#1]}}
\newcommandx*\hilbdimKQ[2][1=,2=]{\mathcal{D}_{K\IfEqCase{#1}{{}{}}[=#1],Q\IfEqCase{#2}{{}{}}[=#2]}}

\newcommand{\poly}{\operatorname{poly}}

\newcommand{\Modep}{{\mathsf{p}}}

\newcommand{\Moden}{{\mathsf{n}}}
\newcommand{\Modem}{{\mathsf{m}}}
\newcommand{\Modek}{{\mathsf{k}}}

\newcommand{\Model}{{\mathsf{l}}}

\newcommand{\Modei}{{\mathsf{i}}}
\newcommand{\Modej}{{\mathsf{j}}}
\newcommand{\mass}{{\mathrm{m}}}
\newcommand{\occ}{{w}}
\newcommand{\occint}{{\textsf{w}}}
\newcommand{\occmax}{W} %
\newcommand{\maxmodes}{I}

\newcommand{\OPa}{a^{\vphantom{\dagger}}}
\newcommand{\OPad}{a^{\dagger}}

\newcommand{\FF}{{\mathcal{F}}}
\newcommand{\interFF}{\widehat{\FF}}
\newcommand{\GG}{{\widetilde{\mathcal{F}}}}
\newcommand{\Dconst}{D^{\vphantom{\dagger}}}

\newcommand{\Kop}{\mathcal{K}}
\newcommandx*{\Hopfree}[1][1=]{\mathcal{H}^{\vphantom{\dagger}}_{\text{free}\IfEqCase{#1}{{}{}}[{,\,#1}]}}
\newcommandx*{\Hopfull}[1][1=]{\mathcal{H}^{\vphantom{\dagger}}_{\text{full}\IfEqCase{#1}{{}{}}[{,\,#1}]}}
\newcommandx*{\Hmatfree}[1][1=]{{H}^{\vphantom{\dagger}}_{\text{free}\IfEqCase{#1}{{}{}}[{,#1}]}}
\newcommandx*{\Hmatfull}[1][1=]{{H}^{\vphantom{\dagger}}_{\text{full}\IfEqCase{#1}{{}{}}[{,#1}]}}
\newcommandx*{\Nopfree}[1][1=]{\mathcal{N}^{\vphantom{\dagger}}_{\text{free}\IfEqCase{#1}{{}{}}[{,#1}]}}
\newcommandx*{\Nopfull}[1][1=]{\mathcal{N}^{\vphantom{\dagger}}_{\text{full}\IfEqCase{#1}{{}{}}[{,#1}]}}
\newcommand{\Efree}{E^{\text{free}}}

\newcommandx*\OPtP[2][1=,2=]{\mathsf{P}_{
\ifx#1#2\empty\else[\fi
\ifx#1\empty\else#1
\ifx#2\empty\else,
\fi
\fi
\ifx#2\empty\else#2
\fi
\ifx#1#2\empty\else]\fi
}}
\newcommandx*\OPtPd[2][1=,2=]{\mathsf{P}^{\dagger}_{
\ifx#1#2\empty\else\fi
\ifx#1\empty\else#1
\ifx#2\empty\else,\,
\fi
\fi
\ifx#2\empty\else#2
\fi
\ifx#1#2\empty\else\fi
}}
\newcommandx*\OPta[2][1=,2=]{\mathsf{a}_{
\ifx#1#2\empty\else[\fi
\ifx#1\empty\else#1
\ifx#2\empty\else,
\fi
\fi
\ifx#2\empty\else#2
\fi
\ifx#1#2\empty\else]\fi
}}
\newcommandx*\OPtad[2][1=,2=]{\mathsf{a}^{\dagger}_{
\ifx#1#2\empty\else[\fi
\ifx#1\empty\else#1
\ifx#2\empty\else,
\fi
\fi
\ifx#2\empty\else#2
\fi
\ifx#1#2\empty\else]\fi
}}
\newcommandx*\OPtA[2][1=,2=]{\mathsf{A}_{
\ifx#1#2\empty\else[\fi
\ifx#1\empty\else#1
\ifx#2\empty\else,
\fi
\fi
\ifx#2\empty\else#2
\fi
\ifx#1#2\empty\else]\fi
}}
\newcommandx*\OPtAd[2][1=,2=]{\mathsf{A}^{\dagger}_{
\ifx#1#2\empty\else[\fi
\ifx#1\empty\else#1
\ifx#2\empty\else,
\fi
\fi
\ifx#2\empty\else#2
\fi
\ifx#1#2\empty\else]\fi
}}

\newcommandx*\Uinit[2][1=,2=]{U_{[{\IfEqCase{#1}{{}{}}#1},{\IfEqCase{#2}{{}{}}#2}]}}
\newcommandx*\Uinitd[2][1=,2=]{U^\dagger_{[{\IfEqCase{#1}{{}{}}#1},{\IfEqCase{#2}{{}{}}#2}]}}

\newcommand{\collective}{{\xi}}
\newcommand{\intercollective}{\widehat{\hspace{0.5pt}\collective}}

\newcommandx*\bound[3][,2=,3=]{#1_{K
\IfEqCase{#2}{{}{}}[,\,#2]}^{
\IfEqCase{#3}{{}{}}[(#3)]
}}

\DeclareMathAlphabet{\mathdutchcal}{U}{dutchcal}{m}{n}

\newcolumntype{C}[1]{>{\centering\let\newline\\\arraybackslash\hspace{0pt}}m{#1}}

\newcommand{\be}{\begin{equation}}
\newcommand{\ee}{\end{equation}}
\newcommand{\bp}{\begin{pmatrix}}
\newcommand{\ep}{\end{pmatrix}}
\newcommand{\ben}{\begin{enumerate}}
\newcommand{\een}{\end{enumerate}}

\def\mystrut(#1,#2){\vrule height #1pt depth #2pt width 0pt}

\newcommandx*{\chargeop}[1][1=]{\mathcal{Q}^
{\IfEqCase{#1}{{}{}}[{(#1)}]}
}
\newcommandx*{\Modecollective}[1][1=]{\upxi^{
\IfEqCase{#1}{{}{}}[{(#1)}]
}
}
\newcommand{\Modeset}{\mathfrak{S}}
\newcommandx*{\statecharge}[1][1=]{{\Upxi^{
\IfEqCase{#1}{{}{}}[{(#1)}]
}}}
\newcommandx*{\FockAllarge}[1][1=]{\Upxi^{
\ifx#1\empty\else#1\fi
}
}

\newcommand{\FockNModes}{J}

\let\oldFootnote\footnote
\newcommand\nextToken\relax
\renewcommand\footnote[1]{\oldFootnote{#1}\futurelet\nextToken\isFootnote}
\newcommand\isFootnote{\ifx\footnote\nextToken\textsuperscript{,}\fi}

\usepackage[shortcuts]{extdash}

\newcommand*\standardbin{+}

\makeatletter

\newcommand{\doublewidetilde}[1]{{%
  \mathpalette\double@widetilde{#1}%
}}
\newcommand{\double@widetilde}[2]{%
  \sbox\z@{$\m@th#1\widetilde{#2}$}%
  \ht\z@=.9\ht\z@
  \widetilde{\box\z@}%
}

\patchcmd{\env@cases}%
  {\quad}%
  {{ \ }}%
  {}{}%

\newcommand*\tabularbin[1]{%
  \mathbin{\mathpalette{\@tabularsym\standardbin}{#1}}%
}

\newcommand*\@tabularsym[3]{%
  \setbox\z@\hbox{$#2#1\m@th$}%
  \hbox to\wd\z@{\hss$#2#3\m@th$\hss}%
}

\newcommand*{\myrulefill}[3][]{%
  \makebox[#2]{#1%
    \leaders\hrule height \dimexpr.5ex+.2pt\relax depth \dimexpr -.5ex+.2pt\relax \hfill%
    \enskip{#3}\enskip%
    \leaders\hrule height \dimexpr.5ex+.2pt\relax depth \dimexpr -.5ex+.2pt\relax \hfill\kern0pt}%
}

\usetikzlibrary{arrows,positioning}
\tikzset{
    >=stealth',
    punkt/.style={
           rectangle,
           rounded corners,
           draw=black, very thick,
           text width=6.5em,
           minimum height=2em,
           text centered},
    pil/.style={
           ->,
           thick,
           shorten <=2pt,
           shorten >=2pt,}
}

\renewcommand\@makecaption[2]{%
  \par
  \vskip\abovecaptionskip
  \begingroup
   \small\rmfamily
    \begingroup
     \samepage
     \flushing
     \let\footnote\@footnotemark@gobble
     \@make@capt@title{#1}{#2}\par
    \endgroup
  \endgroup
  \vskip\belowcaptionskip
}

\def\firstellip{(1.1, -0.3) ellipse [x radius=2.6cm, y radius=1.5cm, rotate=40]}
\def\secondellip{(-1.1, -0.3) ellipse [x radius=2.6cm, y radius=1.5cm, rotate=-40]}
\def\thirdellip{(0, .67) ellipse [x radius=3.8cm, y radius=3.cm]}

\newcommand{\vac}{|{\mathrm{vac}\rangle}}
\newcommand{\vvac}{|{\widetilde{\mathrm{vac}}\rangle}}
\newcommandx*\GSfree[1][1=]{\bigl|{\mathrm{GS},\,{\ifx#1\empty\else#1\fi}\bigr\rangle}_{\mathrm{free}}}
\newcommandx*\nfree[3][1,2,3=]{\bigl|{#2^{\IfEqCase{#3}{{}{}}[#3]}\bigr\rangle}_{\text{free},#1}}
\newcommandx*\snfree[3][1,2,3=]{|{#2^{\IfEqCase{#3}{{}{}}[#3]}\rangle}_{\text{free},#1}}
\newcommandx*\GSfull[1][1=]{\bigl|{\mathrm{GS},\,{\ifx#1\empty\else#1\fi}\bigr\rangle}_{\mathrm{full}}}
\newcommandx*\nfull[3][1,2,3=]{\bigl|{[#1,#2]^{\IfEqCase{#3}{{}{}}[#3]}\bigr\rangle}}
\newcommandx*{\W}[1][1=]{U^{\vphantom{\dagger}}_{\ifx#1\empty\else#1\fi}}
\newcommandx*{\Wd}[1][1=]{U^\dagger_{\ifx#1\empty\else#1\fi}}
\newcommand{\VV}[1][1=]{V^{\vphantom{\dagger}}_{#1}}

\newcommandx*{\Wm}[1][1=]{\mathtt{U}^{\vphantom{\dagger}}_{\ifx#1\empty\else#1\fi}}
\newcommandx*{\Wmd}[1][1=]{\mathtt{U}^\dagger_{\ifx#1\empty\else#1\fi}}
\newcommandx*{\vm}[1][1=]{\mathtt{V}^{\vphantom{\dagger}}_{\ifx#1\empty\else#1\fi}}
\newcommandx*{\vmd}[1][1=]{\mathtt{V}^\dagger_{\ifx#1\empty\else#1\fi}}

\newcommandx*{\Fock}[3][1=,2=,3=]{{\sket{{\FF^{
\ifx#1\empty\else(#1)
\ifx#2\empty\else,\fi
\fi
\ifx#2\empty\else\{#2\}\fi
}
_{
\ifx#3\empty\else#3\fi
}
}}}}
\newcommandx*{\FFock}[3][1=,2=,3=]{{\sket{{\GG^{
\ifx#1\empty\else(#1)
\ifx#2\empty\else,\fi
\fi
\ifx#2\empty\else\{#2\}\fi
}
_{
\ifx#3\empty\else#3\fi
}
}}}}

\newcommandx*{\FockAll}[1][1=]{{\mathfrak{F}^{\ifx#1\empty\else\{#1\}\fi}}}
\newcommandx*{\FFockAll}[1][1=]{{\widetilde{\mathfrak{F}}^{\ifx#1\empty\else\{#1\}\fi}}}

\newcommandx*{\Wop}[1][1=]{\mathcal{U}^{\vphantom{\dagger}}_{\ifx#1\empty\else#1\fi}}
\newcommandx*{\Wopd}[1][1=]{\mathcal{U}^{{\dagger}}_{\ifx#1\empty\else#1\fi}}
\newcommandx*{\Vop}[2][1=,2=]{\mathcal{V}^{{{\IfEqCase{#2}{{}{\vphantom{\dagger}}}[{\, #2}]}}}_{\ifx#1\empty\else#1\fi}}
\newcommandx*{\Vopd}[1][1=]{\mathcal{V}^{{\dagger}}_{\ifx#1\empty\else#1\fi}}
\newcommand {\oprelat} [2] {o^{\vphantom{\dagger}}_{#1 | #2}}
\newcommand {\oprelatd} [2] {o^{\dagger}_{#1 | #2}}

\newcommandx*{\gsop}[2][1=,2=]{{}^{\mathsf{(#2)}}\mathsf{g}^\dagger_{\mathsf{#1}}}

\newcommandx*{\OOPad}[1][1=]{{}^{{\mathsf{#1}}}{\mathsf{a}}^{\dagger}}
\newcommand{\OOPA}{\mathsf{A}^{\vphantom{\dagger}}}
\newcommand{\OOPAd}{\mathsf{A}^{\dagger}}

\makeatother

\makeatletter
\let\save@mathaccent\mathaccent
\newcommand*\if@single[3]{%
  \setbox0\hbox{${\mathaccent"0362{#1}}^H$}%
  \setbox2\hbox{${\mathaccent"0362{\kern0pt#1}}^H$}%
  \ifdim\ht0=\ht2 #3\else #2\fi
  }
\newcommand*\rel@kern[1]{\kern#1\dimexpr\macc@kerna}
\newcommand*\wideaccent[2]{\@ifnextchar^{{\wide@accent{#1}{#2}{0}}}{\wide@accent{#1}{#2}{1}}}
\newcommand*\wide@accent[3]{\if@single{#2}{\wide@accent@{#1}{#2}{#3}{1}}{\wide@accent@{#1}{#2}{#3}{2}}}
\newcommand*\wide@accent@[4]{%
  \begingroup
  \def\mathaccent##1##2{%
    \let\mathaccent\save@mathaccent
    \if#42 \let\macc@nucleus\first@char \fi
    \setbox\z@\hbox{$\macc@style{\macc@nucleus}_{}$}%
    \setbox\tw@\hbox{$\macc@style{\macc@nucleus}{}_{}$}%
    \dimen@\wd\tw@
    \advance\dimen@-\wd\z@
    \divide\dimen@ 3
    \@tempdima\wd\tw@
    \advance\@tempdima-\scriptspace
    \divide\@tempdima 10
    \advance\dimen@-\@tempdima
    \ifdim\dimen@>\z@ \dimen@0pt\fi
    \rel@kern{0.6}\kern-\dimen@
    \if#41
      #1{\rel@kern{-0.6}\kern\dimen@\macc@nucleus\rel@kern{0.4}\kern\dimen@}%
      \advance\dimen@0.4\dimexpr\macc@kerna
      \let\final@kern#3%
      \ifdim\dimen@<\z@ \let\final@kern1\fi
      \if\final@kern1 \kern-\dimen@\fi
    \else
      #1{\rel@kern{-0.6}\kern\dimen@#2}%
    \fi
  }%
  \macc@depth\@ne
  \let\math@bgroup\@empty \let\math@egroup\macc@set@skewchar
  \mathsurround\z@ \frozen@everymath{\mathgroup\macc@group\relax}%
  \macc@set@skewchar\relax
  \let\mathaccentV\macc@nested@a
  \if#41
    \macc@nested@a\relax111{#2}%
  \else
    \def\gobble@till@marker##1\endmarker{}%
    \futurelet\first@char\gobble@till@marker#2\endmarker
    \ifcat\noexpand\first@char A\else
      \def\first@char{}%
    \fi
    \macc@nested@a\relax111{\first@char}%
  \fi
  \endgroup
}
\makeatother

\newcommand\widebar{\wideaccent\overline}

\newcommand{\Chargeset}{\widebar{\mathfrak{S}}}

\newcommand{\Hflow}{\mathcal{H}^{\vphantom{\dagger}}}
\newcommand{\Uflow}{\mathcal{U}^{\vphantom{\dagger}}}
\newcommand{\Udflow}{\mathcal{U}^{{\dagger}}}
\newcommand{\Gflow}{\mathcal{G}^{\vphantom{\dagger}}}



\begin{document}

\title{Simulating Scattering of Composite Particles}

\author{Michael Kreshchuk}
\affiliation{Physics Division, Lawrence Berkeley National Laboratory, Berkeley, California 94720, USA}
\affiliation{Department of Physics and Astronomy, Tufts University, Medford, MA 02155, USA}
\author{James P. Vary}
\affiliation{Department of Physics and Astronomy, Iowa State University, Ames, IA 50011, USA}
\author{Peter J. Love}
\affiliation{Department of Physics and Astronomy, Tufts University, Medford, MA 02155, USA}
\affiliation{Computational Science Initiative, Brookhaven National Laboratory, Upton, NY 11973, USA }

\begin{abstract}
We develop a non-perturbative approach to simulating scattering on classical and quantum computers, in which the initial and final states contain a fixed number of composite particles.
The construction is designed to mimic a particle collision, wherein two composite particles are brought in contact.
The initial states are assembled via consecutive application of operators creating eigenstates of the interacting theory from vacuum.
These operators are defined with the aid of the M{\o}ller wave operator, which can be constructed using such methods as adiabatic state preparation or double commutator flow equation.

The approach
is well-suited for studying strongly coupled systems in both relativistic and non-relativistic settings.
For relativistic systems, we employ the language of light-front quantization, which has been previously used for studying the properties of individual bound states, as well as for simulating their scattering in external fields, and is now adopted to the studies of scattering of bound state systems.

For simulations on classical computers, we describe an algorithm for calculating exact (in the sense of a given discretized theory) scattering probabilities, which has cost (memory and time) exponential in momentum grid size.
Such calculations may be interesting in their own right and can be used for benchmarking results of a quantum simulation algorithm, which is the main application of the developed framework.
We illustrate our ideas with an application to the $\phi^4$ theory in $1+1\mathrm{D}$.

\end{abstract}

\maketitle

\section{Introduction}
Calculation of scattering probabilities is one of the primary tasks of quantum field theory (QFT).
Most analytic approaches describe scattering using the S-matrix, a unitary operator connecting the states of incoming and outgoing particles, which are considered asymptotically free.
S-matrix theory has proven enormously successful in the studies of weakly coupled quantum field theories, such as quantum electrodynamics or high-energy quantum chromodynamics~\cite{Peskin:1995ev}.

Most proposals for \emph{ab initio} simulation of scattering in QFT rely either on significant reduction of problem complexity (i.e., truncation of many-body Hilbert space to a small number of particles and/or classical description of fields~\cite{PhysRevD.88.065014,PhysRevC.97.064620,PhysRevA.104.012611,Lei:2022nsk}) or on the usage of quantum hardware~\cite{2011arXiv1112.4833J,preskill1,preskill2,garcia2015fermion,chiesa2019quantum,gustafson2019quantum,Surace_2021,atas20212,turco2023towards,barata2023quantum,Belyansky:2023rgh}, with the latter being the primary application of results of this work.

Since the early days of quantum computation~\cite{feynman1982,feynman1986quantum}, quantum simulation has been recognized as one of its primary applications, and was considered as a way of achieving
a long-time dream of quantum physicists~--- solving quantum field theory non-perturbatively.
Currently, quantum simulation is the only
proposal offering a way of making calculations in general quantum many-body systems and field theories, that requires computational resources growing polynomially with the problem size and precision~\cite{lloyd1996universal,zalka1998simulating,wiesner1996simulations,boghosian1997quantum,meyer1996quantum,aspuru2005simulated,wu2002polynomial,2011arXiv1112.4833J,preskill1,preskill2}.
While certain properties of relativistic bound states can be calculated on near-term devices with limited resources~\cite{Kreshchuk:2020kcz,Kreshchuk:2020aiq,qian2022solving}, simulation of scattering is likely to require quantum computers functioning in the fault-tolerant regime~\cite{gaitan2008quantum,gottesman2009introduction,lidar2013quantum,litinski2019game}.

Most existing approaches~\cite{2011arXiv1112.4833J,preskill1,preskill2,gustafson2019quantum,Surace_2021,atas20212,turco2023towards} to the quantum simulation of scattering comprise the following steps: (a)~non-trivial preparation of the vacuum state; (b)~preparation of wave packets;\footnote{While our approach does not involve the use of wave packets, it is useful to recall why this step is typically considered necessary.
The argument is as follows: since a multi-particle state in the free theory is an eigenstate of the free Hamiltonian, an adiabatic interaction turn-on  would transform this state into an eigenstate of the full Hamiltonian, which evolves by phase multiplication. Non-trivial time evolution therefore requires wavepackets: coherent superpositions of eigenstates, localized around particular expectation values of position and momentum.
} (c)~adiabatic interaction turn-on and turn-off; (d)~measurement.
In this work, we propose a paradigm for simulation of scattering, in which: (a)~the initial and final scattering states are defined as states of a fixed number of composite particles in the interacting theory; (b)~the many-particle states belong to the theory with momentum cut-offs higher than those in the theories describing the states of individual particles.
That is, the addition of new particles into the system requires the addition of higher momentum modes, upon which the Hamiltonian operator can act.
Due to the increase of the total momentum of the system, states containing multiple composite particles are not eigenstates of the combined system, and, therefore, undergo non-trivial time evolution.

Our approach combines ideas from non-relativistic many-body physics~\cite{szabo1996modern,doi:10.1063/1.477094,C9CP06376E}, quantum field theory~\cite{pauli1,pauli2,phi4in2d,Brodsky:1997de,varybasis,PhysRevD.101.076016,kuang2022all} and quantum simulation~\cite{preskill1,atas20212,Surace_2021}.
Furthermore, it is based upon the second-quantized formulation, and so allows one to employ appropriate basis sets to efficiently describe systems
involving
localized bound states~\cite{varybasis}.

In application to high-energy physics, our work can be considered as an extension of the time-dependent Basis Light Front Quantization (tBLFQ) which has been successfully used for a number of scattering applications to date~\cite{PhysRevD.88.065014,PhysRevC.97.064620,PhysRevA.104.012611,Lei:2022nsk,barata2023quantum}.
In tBLFQ, the initial and final states of the system are typically chosen to be the eigenstates of some interacting Hamiltonian.
The non-trivial evolution of such eigenstates in time is then achieved by coupling the original system to an external field.
This, for example, allows one to naturally model the scattering of light particles on heavy nuclei~\cite{PhysRevD.101.076016}.
We generalize this construction to include situations in which scattering is caused by the interacting quantum field itself, as in the cases when beams of particles of comparable mass are scattered.

\begin{figure}
    \centering
    \begin{tikzpicture}[
        ball/.style={circle, shading=ball, ball color=black!15, minimum size=9mm},
    ]

    \node (B) [ball] at (-2,1) {$\sket{\widetilde{\bm{p}}_1}^{(\Lambda_1)}$};
    \node (D) [ball] at (2,1) {$\sket{\widetilde{\bm{p}}_2}^{(\Lambda_2)}$};

    \node (E) [ball] at (0,-1) {$\sket{\widetilde{\bm{p}}_1,\widetilde{\bm{p}}_2}^{(\Lambda_1+\Lambda_2)}$};

    \draw [-Latex] (B) -- (E);
    \draw [-Latex] (D) -- (E);

\end{tikzpicture}
    \caption{Two particles of momenta $\bm{p}_1$ and $\bm{p}_2$ are described as the interacting Hamiltonian eigenstates ${\sket{\widetilde{\bm{p}}_1}^{(\Lambda_1)}=\OOPAd_{\Lambda_1,{\bm{p}}_1}\vac}$ and ${\sket{\widetilde{\bm{p}}_2}^{(\Lambda_2)}=\OOPAd_{{\Lambda_2,\,\bm{p}}_2}\vac}$, whose mode momentum cutoffs are $\Lambda_1$ and $\Lambda_2$.
    The combined state ${\sket{\widetilde{\bm{p}}_1,\widetilde{\bm{p}}_2}^{(\Lambda_1+\Lambda_2)}=\OOPAd_{\Lambda_1,{\bm{p}}_1}\OOPAd_{\Lambda_2,{\bm{p}}_2}\vac}$ contains modes of momenta up to ${\Lambda_1+\Lambda_2}$.
    This state is, generally, \emph{not} an eigenstate of the Hamiltonian operator acting on momentum modes up to ${\Lambda_1+\Lambda_2}$, and undergoes non-trivial time evolution.}
    \label{fig:particle_cartoon}
\end{figure}

In~\Cref{sec:main_idea}, we present the main ideas of our approach, while leaving a more general discussion to~\Cref{sec:definitions}.
In~\Cref{sec:dlcq} we illustrate those using as an example the formulation of the $\phi^4$ theory in $1+1\mathrm{D}$ obtained within the Discretized Light-Cone Quantization framework~\cite{phi4in2d}.
In~\Cref{sec:scatter_prob}, we discuss a detailed implementation of our approach via a purely classical calculation.
This will require resources exponential in the momentum grid size, and will give the exact solution to the problem.
Here and in what follows, \emph{exact solution} refers to an solution of the chosen discretized model approximating the original continuous theory.
As such, we do not address the effects arising at the stage of discretizing the continuous theory.
One way to approach the ``exact'' solution in the sense of the continuous theory amounts to extrapolating simulation results to infinite values of cutoffs~\cite{Vary:2021cbh}.
In~\Cref{sec:scat_quant} we describe an efficient quantum simulation algorithm, which is the main motivation for introducing the new framework.

\section{States of multiple composite particles\label{sec:main_idea}}

In this Section, we introduce the main conceptual ideas of our method.
For clarity, we will make certain simplifications (e.g. adopt the momentum basis for single-particle states and ignore other quantum numbers) while deferring a more general discussion to~\Cref{sec:definitions}.

Consider a non-interacting theory described by the Hamiltonian operator $\Hopfree[\Lambda]$, provided that $\Hopfree[\Lambda]$ can be written in terms of creation and annihilation operators, and modes of momentum up to $\Lambda$ are included.
In such a theory, the single-particle states of momentum $\abs{\bm{p}}\leq\Lambda$ are created from the vacuum state as
\begin{equation}
    \label{eq:freefromvac}
    {\sket{\bm{p}}}^{(\Lambda)} = \OPad_{\Lambda,\,\bm{p}} \vac \, .
\end{equation}
Having in mind applications of our formalism to non-relativistic many-body theory and Light-Front (LF) QFT~\cite{pauli1,pauli2,phi4in2d,Brodsky:1997de,varybasis} (see also~\cref{foot:zero} below), here and in what follows we assume the uniqueness of the vacuum state, which is the same for both free and interacting theories~\cite{fetter2003quantum,BRODSKY1998299}.
The notation ${\sket{\bm{p}}}^{(\Lambda)}$ in~\cref{eq:freefromvac} was introduced solely for uniformity purposes, as ${\OPad_{\Lambda_1,\,\bm{p}}\equiv \OPad_{\Lambda_2,\,\bm{p}}\equiv\OPad_{\bm{p}}}$ and ${{\sket{\bm{p}}}^{(\Lambda_1)}\equiv{\sket{\bm{p}}}^{(\Lambda_2)}\equiv{\sket{\bm{p}}}^{}}$ (where it is assumed that %
$\abs{\bm{p}}\leq\min{(\Lambda_1,\Lambda_2)})$.

Note that in order to include states of the form ${\OPad_{\Lambda_1,\,\bm{p}_1} \OPad_{\Lambda_2,\,\bm{p}_2} \vac}$, one generally has to consider a theory with the cutoff being at least ${\Lambda_1+\Lambda_2}$:
\begin{equation}
\label{eq:manybfree}
    \sket{\bm{p}_1,\bm{p}_2}^{(\Lambda_1+\Lambda_2)}
    =
    \OPad_{\Lambda_1,\,\bm{p}_1}
    \OPad_{\Lambda_2,\,\bm{p}_2} \vac \, .
\end{equation}
In this way one can always define the action of ${\OPad_{\Lambda,\,\bm{p}}}$ on any Fock state~--- in accord with its action in the theory with ${\Lambda=\infty}$, which includes Fock states of arbitrary momentum.

In analogy with~\cref{eq:freefromvac}, in the interacting theory  described by the Hamiltonian $\Hopfull[\Lambda]$, we introduce the second-quantized operator $\OOPAd_{{\Lambda,\,\bm{p}}}$ creating a single-particle state of momentum $\bm{p}$ (in the sense of the interacting theory) from vacuum:
\begin{equation}
\label{eq:pavac}
{\sket{\widetilde{\bm{p}}}^{(\Lambda)}=\OOPAd_{{\Lambda,\,\bm{p}}}\vac} \, .
\end{equation}

One possible way of constructing $\OOPAd_{\Lambda,\,\bm{p}}$ amounts to writing ${\sket{\widetilde{\bm{p}}}^{(\Lambda)}}$ as a polynomial $\OPtPd[\Lambda][\bm{p}]$ in the free creation operators $\OPad_{\Lambda,\,\bm{p}}$ of momenta up to $\Lambda$, acting on the vacuum state,
\begin{equation}
\label{eq:defP}
\begin{gathered}
    \sket{\widetilde{\bm{p}}}^{(\Lambda)}= \OPtPd[\Lambda][\bm{p}] \vac \,,\\
    \begin{alignedat}9
    \OPtPd[\Lambda][\bm{p}] &= \sum_n \int^\Lambda \d{ \bm{p}_1}\ldots \d{ \bm{p}_n}
    \\&\times
    f^{(n)}(\bm{p_1},\ldots,\bm{p_n}) \prod_{j=1}^n \OPad_{\Lambda,\,\bm{p}_n} \,,
    \end{alignedat}
\end{gathered}
\end{equation}
and then defining ${\OOPAd_{\Lambda,\,\bm{p}} \equiv \OPtPd[\Lambda][\bm{p}] }$.
Note, however, that since~$\OPtPd[\Lambda][\bm{p}]$ only carries information on the state $\sket{\widetilde{\bm{p}}}^{(\Lambda)}$ (in the form of how the latter can be prepared from the vacuum), its action on any other state in the Fock space is not, generally, particularly sensible.
Nevertheless, in certain cases it may serve as an approximate version of the creation operator defined below.

Alternatively, one can define $\OOPAd_{{\Lambda,\,\bm{p}}}$ by unitarily rotating the operator $\OPad_{\Lambda,\,\bm{p}}$:
\begin{equation}
    \label{eq:creatop}
    \OOPAd_{\Lambda,{\bm{p}}} = \Wop[\Lambda] \OPad_{\Lambda,\,\bm{p}}\, \Wopd[\Lambda] \, ,
\end{equation}
with $\Wop[\Lambda]$ being a transformation relating the bases of the free and interacting theories.
In this case, using the language of spectral theory, we shall refer to $\Wop[\Lambda]$ as M{\o}ller \emph{wave operator}~\cite{newton2013scattering,kato1966wave,nguyen1985review,evangelista2014driven}.
Definition~\eqref{eq:creatop} is consistent with~\cref{eq:pavac},
\begin{equation}
\label{eq:justif}
\begin{alignedat}{8}
    \OOPAd_{\Lambda,{\bm{p}}} \vac
    &=
    \Wop[\Lambda] \OPad_{\Lambda,\,\bm{p}} \Wopd[\Lambda]
    \vac
    \\
    &=
    \Wop[\Lambda]
    \OPad_{\Lambda,\,\bm{p}}
    \vac
    =\sket{\widetilde{\bm{p}}}^{(\Lambda)}
    \, ,
\end{alignedat}
\end{equation}
given that $\Wop[\Lambda]$ acts on the vacuum and single-particle states as follows:\footnote{\label{footnote:vacua}Equations~\eqref{eq:justif} and~\eqref{eq:vac_inv} can be generalized to theories with non-trivial interacting vacua as
 $\OOPAd_{\Lambda,{\bm{p}}} \vvac^{(\Lambda)} = \Wop[\Lambda] \OPad_{\Lambda,\,\bm{p}} \Wopd[\Lambda]
\Wop[\Lambda]
    \vac$ and $\Wop[\Lambda] \vac = \vvac^{(\Lambda)}$, where $\vvac^{(\Lambda)}$ is the vacuum state in the interacting theory with cutoff $\Lambda$.
    In such cases, one would also need to distinguish vacua of interacting theories with different cutoffs.
}
\begin{alignat}9
    \label{eq:vac_inv}
    &\Wop[\Lambda] \vac &&= \vac \, &&,\\
    &\Wop[\Lambda]
    \sket{\bm{p}}^{(\Lambda)}
    &&=\sket{\widetilde{\bm{p}}}^{(\Lambda)}\,&&.
\end{alignat}

If $\Wop[\Lambda]$ respects the symmetries of the theory, definition~\eqref{eq:creatop} implies that $\OOPAd_{\Lambda,{\bm{p}}}$ and $\OPad_{\Lambda,\,\bm{p}}$ create states with the same values of charges.
For an operator creating a state in the interacting theory whose quantum numbers differ from those of the free particles (such as proton), \cref{eq:creatop} has to be generalized by replacing $\OPad_{\Lambda,\,\bm{p}}$ with an operator creating the ground state of the free theory in the sector with corresponding charges, see~\Cref{sec:definitions}.

Definition~\eqref{eq:creatop} is closely related to the notion of \emph{effective particles} in QFT~\cite{glazek1993renormalization,glazek1994perturbative,glazek2012perturbative,glazek2012renormalization,glazek2013fermion,Serafin:2019vuk,glazek2021elementary,glazek2017renormalized,glazek2017effective,gomez2017asymptotic}.
Unlike $\OPtPd[\Lambda][\bm{p}]$, the wave operator $\Wop[\Lambda]$ carries information about the entire spectrum of the system, up to momentum cutoff $\Lambda$.
Unless otherwise specified, in the remainder of the paper we shall, therefore, assume the usage of definition~\eqref{eq:creatop}.

Note that, in contrast with the free theory where ${\OPad_{\Lambda,\,\bm{p}}}$ simply increases the occupancy of mode $\bm{p}$ in a Fock state, in the interacting theory $\OOPAd_{\Lambda,\,\bm{p}}$ may act on all modes of momentum up to $\Lambda$, which implies that, generally, ${\OOPAd_{\Lambda_1,\,\bm{p}}\neq\OOPAd_{\Lambda_2,\,\bm{p}}}$ and ${\sket{\widetilde{\bm{p}}}^{(\Lambda_1)}\neq\sket{\widetilde{\bm{p}}}^{(\Lambda_2)}}$.

Following the analogy with the free theory, we define states with a fixed number of particles in the interacting theory by successive application of creation operators~\eqref{eq:creatop} to vacuum:
\begin{equation}
    \label{eq:manybint}
    \OOPAd_{\Lambda_1,\,\bm{p}_1} \OOPAd_{\Lambda_2,\,\bm{p}_2} \ldots \vac \, ,
\end{equation}
which mimics the definition~\eqref{eq:manybfree}.
Here we set aside the detailed issue of normalization of state~\eqref{eq:manybint}.

Note that whether the operators $\OOPAd_{\Lambda_j,\,\bm{p}_j}$ are defined as in~\cref{eq:defP} or as in~\cref{eq:creatop},
the states in~\eqref{eq:manybint} are not the eigenstates of $\Hopfull[\Lambda_1+\Lambda_2+\ldots]$, since neither construction assumes solving $\Hopfull[\Lambda_1+\Lambda_2+\ldots]$.
Therefore, the time evolution of states of the form~\eqref{eq:manybint} under the action of $\Hopfull[\Lambda_1+\Lambda_2+\ldots]$ is non-trivial, and involves modes of higher momentum, which are absent in the one-particle states but present in the many-particle state~\eqref{eq:manybint}.
Intuitively, this picture reflects the fact that bringing new particles into the system increases its total momentum and entails inclusion of higher-momentum modes in the interaction, see~\Cref{fig:particle_cartoon}.
The so-defined states are then evolved with time and used for measurement.

Operators creating particles in the interacting system are often chosen to act on disjoint sets of modes.
If spatial discretization is used, this can be achieved by imposing a finite spatial extent of particle wavefunctions and assuming large separation in the coordinate space between them~\cite{preskill1,Surace_2021,atas20212,kuang2022all}.
In the momentum basis, a similar situation is encountered if the center-of-mass momenta of composite particles are large compared to relative momenta of their constituents.
While definitions~\eqref{eq:defP} and~\eqref{eq:creatop} are not equivalent in the sense of their action on arbitrary Fock states, both lead to the two-particle wavefunction ${\OOPAd_{\Lambda_1,\,\bm{p}_1}\OOPAd_{\Lambda_2,\,\bm{p}_2}\vac}$ being a tensor product of ${\OOPAd_{\Lambda_1,\,\bm{p}_1}\vac}$ and ${\OOPAd_{\Lambda_2,\,\bm{p}_2}\vac}$ when ${\OOPAd_{\Lambda_1,\,\bm{p}_1}}$ and ${\OOPAd_{\Lambda_2,\,\bm{p}_2}}$ act on disjoint sets of free particle modes~\cite{simenel2008microscopic}.

The situation becomes more complicated if the sets of modes, upon which $\OOPAd_{\Lambda_1,\,\bm{p}_1}$ and $\OOPAd_{\Lambda_2,\,\bm{p}_2}$ act, \emph{do} overlap~--- an example of such a scenario is considered below in~\Cref{sec:dlcq,sec:scatter_prob}, in the context of LF quantization.\footnote{Considering overlapping wavefunction of hadrons at the stage of initial state preparation may be necessary not only in the LF formulation but for equal-time gauge theories as well~\cite{maiani1990final}.}
In this case, definitions~\eqref{eq:defP} and~\eqref{eq:creatop} may lead to significantly different forms of the state~${\OOPAd_{\Lambda_1,\,\bm{p}_1}\OOPAd_{\Lambda_2,\,\bm{p}_2}\vac}$, neither of which would be a tensor product of ${\OOPAd_{\Lambda_1,\,\bm{p}_1}\vac}$ and ${\OOPAd_{\Lambda_2,\,\bm{p}_2}\vac}$; see the discussion at the end of this section in~\cref{item:noncomm}.

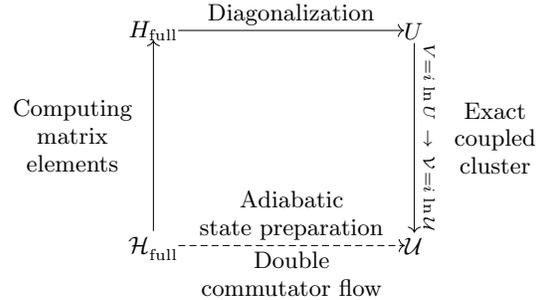
\begin{figure}
\tikzset{
    labl/.style={anchor=south, rotate=270, inner sep=.5mm}
}
\[
\begin{tikzcd}[row sep=2.5cm, column sep = 3.cm]
H_{\text{full}} \arrow{r}{\parbox{2.5cm}{Diagonalization}}  & \W \arrow[d,"{\parbox{2cm}{Exact\\coupled\\cluster\\
}
}","V=\iu\ln U \  \to \ \Vop = \iu \ln \Wop"' labl] \\%
\Hopfull \arrow{u}{\parbox{2cm}{Computing matrix\\elements}} \arrow[r,dashed,"\parbox{3cm}{Adiabatic\\state preparation}","\parbox{2.5cm}{Double commutator flow}"']& \Wop
\end{tikzcd}
\]
    \caption{Approaches to constructing the \emph{wave operator}~\cite{kato1966wave,nguyen1985review,evangelista2014driven} $\Wop$ relating the eigenbases of the free $\Hopfree$ and full $\Hopfull$ Hamiltonian operators.
    Following the solid lines renders the exact solution to the problem: $\Wop$ can be obtained by diagonalizing the matrix $H_{\text{full}}$ of $\Hopfull$, forming the \emph{modal matrix} $U$ comprised of the eigenvectors of $H_{\text{full}}$, and finding such a cluster operator ${\Vop=\iu\ln\Wop}$ that its matrix elements are identical to those of ${V = \iu \ln U}$ (see~\Cref{app:ucc} for details).
    Explicitly using the matrix $H_{\text{full}}$ results in a procedure whose cost is polynomial in the Hilbert space dimension and, therefore, exponential in momentum cutoffs.
    The dashed line corresponds to approximate solutions to the problem: constructing $\Wop$ by employing adiabatic state preparation or solving the double commutator flow equation~\cite{glazek1993renormalization,glazek1994perturbative,wilson1994nonperturbative,wegner1994flow}. Implementing those on a quantum computer~\cite{Aharonov:2003:AQS:780542.780546,aspuru2005simulated,du2010nmr,2011arXiv1112.4833J,veis2014adiabatic,barends2016digitized,wan2020fast,sugisaki2022adiabatic,ciavarella2023state,gluza2022double} may be efficient, i.e., have cost polynomial in momentum cutoffs.
    }
    \label{fig:u}
\end{figure}

Several ways of constructing the wave operator $\Wop$ can be considered (for brevity, we suppress the dependence of $\Wop$ on $\Lambda$ in the remainder of the section), see~\Cref{fig:u}:
\begin{itemize}[ leftmargin = 15pt ]
\item Within the unitary coupled cluster (UCC) method, one seeks for $\Wop$ in the form of exponential of a Hermitian polynomial $\Vop$ in creation and annihilation operators:
\begin{equation}
\label{eq:eucc}
\begin{gathered}
\Wop = \me^{-\iu \Vop}\,,\\
\Vop = \poly(\OPa_{\bm{p}},\OPad_{\bm{p}})\,,
\quad
\Vop = \Vopd\,.
\end{gathered}
\end{equation}

When the UCC construction is used for approximating low-lying states in the spectrum of $\Hopfull$, $\Vop$ may include only low powers of creation and annihilation operators, while the coefficients in the polynomial are found by means of a heuristic procedure~\cite{shen2017quantum,Romero_2018,filip2020stochastic}.
However, in order for $\Wop$ to implement the exact rotation between the eigenbases of $\Hopfree$ and $\Hopfull$, the cluster operator $\Vop$ has to include a significantly larger number of terms.
The procedure for finding those is discussed in~\Cref{app:ucc}, and will be referred to as \emph{exact} UCC.

\item Implementing $\Wop[\Lambda]$ presents a new challenge to quantum simulation methods. Adiabatic state preparation uses quantum simulation of a time-dependent Hamiltonian to prepare the ground state of some target Hamiltonian~\cite{Aharonov:2003:AQS:780542.780546,aspuru2005simulated,du2010nmr,2011arXiv1112.4833J,veis2014adiabatic,barends2016digitized,wan2020fast,sugisaki2022adiabatic,ciavarella2023state}. In our case, we first wish to apply $\Wop[\Lambda]$ to $\OPad_{\Lambda,\,\bm{p}}\vac$. The state $\OPad_{\Lambda,\,\bm{p}}\vac$ is the ground state of the single particle sector of the interacting theory. Therefore one application  of $\Wop[\Lambda]$ is a map between ground states of different theories. Time evolution under a time dependent Hamiltonian which interpolates between the initial and final Hamiltonian will obey the adiabatic theorem and hence will drag one ground state into another provided the evolution is slow enough. The maximum speed is set by the gap between the ground and first excited states. This gap may be unknown, but we could proceed using the procedure outlined in~\cite{2011arXiv1112.4833J} in which the gap is estimated as one goes along, and this estimate is used to control the speed of the adiabatic evolution.

However, our goal is to produce multiparticle states of the interacting theory by sequential application of raising operators. The second application of $\OOPAd_{{\Lambda,\,\bm{p}}}$ involves adiabatically turning on $\Wop[\Lambda]$ but this time with a starting state that is neither a ground state nor a superposition of ground states. This is different from the case of adiabatic state preparation of wavepackets, which in the free theory are superpositions of single-particle states, each being a ground state of the free Hamiltonian in the sector of particular momentum.

In principle one could imagine the adiabatic theorem holding for all eigenvalues along the path, however in general this cannot be the case. We expect that $\Wop[\Lambda]$ will have an exponentially large number of eigenvalues, while its generating Hamiltonian will have a polynomially bounded norm. It is therefore in  general impossible to ``squeeze" the whole spectrum into the norm of the Hamiltonian without having two or more levels coming exponentially close to one another.

Moreover, the states to which $\Wop[\Lambda]$ is applied no longer belong to the theory used for defining the action of $\Wop[\Lambda]$ (they, instead, act in some larger Hilbert space, see, e.g.,~\cref{eq:2K} below), which raises the question of what the conditions for implementing $\Wop[\Lambda]$ in such a scenario even are.
To the best of our knowledge, such a question has not been discussed in the literature.

The problem posed for quantum simulation therefore calls for new techniques. One possible avenue is discussed in the next section.

\item $\Uflow$ can also be constructed using the G\l{}azek-Wilson-Wegner (GWW) double commutator flow equation~\cite{glazek1993renormalization,glazek1994perturbative,wilson1994nonperturbative,wegner1994flow,glazek2012perturbative,glazek2012renormalization,glazek2013fermion,Serafin:2019vuk,glazek2021elementary,glazek2017effective,glazek2017renormalized,gomez2017asymptotic,wilson1994nonperturbative,wegner1994flow}.
While in the context of LF QFT, the double commutator flow equation technique is known as the Renormalization Group Procedure for Effective Particles \mbox{(RGPEP)}~\cite{glazek1993renormalization,glazek1994perturbative,glazek2012perturbative,glazek2012renormalization,glazek2013fermion,Serafin:2019vuk,glazek2021elementary,glazek2017effective,glazek2017renormalized,gomez2017asymptotic}, and has been predominantly used for renormalizing continuous theories which were then solved by various numerical techniques~\cite{glazek2012renormalization,glazek2013fermion,Serafin:2019vuk,glazek2021elementary},
recently a number of numerical schemes based on the double commutator flow equation have been suggested~\cite{evangelista2014driven,li2015multireference,li2016towards,hannon2016integral,li2017driven,li2017low,li2018driven,li2019multireference,wang2019analytic,zhang2019improving}, including an implementation for
quantum computers~\cite{gluza2022double}.
Unlike the adiabatic methods, the GWW approach can be used to describe phase transitions~\cite{bartlett2003flow,quito2016localization,kohler2013asymmetry,kohler2013nonequilibrium}, which is crucial in the studies of confinement in quantum chromodynamics~\cite{glazek2017renormalized,gomez2017asymptotic,Serafin:2019vuk}.

In the GWW approach, one defines a family of unitary operators labeled by a continuous parameter ${l\in[0,\infty)}$ which perform a \emph{similarity transformation} of the original Hamiltonian (which includes all the interaction terms) in such a way that increasing the evolution parameter $l$ entails the decrease of non-diagonal terms, with $\Hflow_{\infty}$ ultimately becoming diagonal in the limit~${l\to\infty}$.

The original ${\Hflow_0}$ and transformed $\Hflow_l$ Hamiltonians are related via
\begin{equation}
\Hflow_l = \Udflow_l \Hflow_0\Uflow_l\,,
\end{equation}
where $\Hflow_l$ obeys the flow equation
\begin{equation}
\partial_l^{\vphantom{\dagger}} \Hflow_l
= [\Gflow_l,\Hflow_l]\,
\end{equation}
with the \emph{flow generator} ${\Gflow_l=(\partial_l^{\vphantom{\dagger}}\Uflow_l)\Udflow_l}$ defining the form of~$\Uflow_l$.
In
non-relativistic theories, ${\Gflow_l}$ can be defined as a commutator between the diagonal $\Delta(\Hflow_l)$ and off-diagonal $\sigma(\Hflow_l)$ parts of the Hamiltonian operator:
\begin{equation}
    \Gflow_l = [\Delta(\Hflow_l), \sigma(\Hflow_l)] \,.
\end{equation}
In the LF formulation, this changes to
\begin{equation}
    \label{eq:flowgrel}
    \Gflow_l = [\Delta(\Hflow_l), \tilde{\sigma}(\Hflow_l)] \,,
\end{equation}
where $\tilde{\sigma}(\Hflow_l)$ has additional structure ensuring the Lorentz invariance of the flow~\cite{glazek2012perturbative}.
Other choices of $\Gflow_l$ are possible as well~\cite{glazek2012perturbative}.

\item Preparation of states with multiple composite particles is most easily achieved when the corresponding creation operators act on disjoint sets of degrees of freedom, and the wave function factorizes.
As in such cases the action of operators, creating composite particles in the interacting theory, has to be defined only upon the vacuum state, a wider class of state preparation techniques is readily available in addition to those discussed above, including variational~\cite{Surace_2021,atas20212} and projection-based methods~\cite{ge2019faster,lin2020near,dong2022ground}.

\end{itemize}

Several other remarks are in order regarding the definition~\eqref{eq:manybint} of scattering states.

\begin{enumerate}[label={(\alph*)}, leftmargin = 15pt ]

\item The suggested formalism is well-suited for situations in which the incoming and outgoing particles are the states of strongly interacting systems, such as hadrons or heavy nuclei, which can be approximated neither by the eigenstates of the free Hamiltonian nor by wavepackets.

\item In the relativistic setting, our approach is most naturally applicable to studying systems described in the language of LF quantization~\cite{pauli1, pauli2, phi4in2d, Brodsky:1997de, varybasis} which recasts relativistic many-body problems in a form that is strikingly similar to non-relativistic many-body theory.
For example, the unique LF vacuum state of the free theory coincides with that one of the interacting theory.\footnote{\label{foot:zero}Certain light-front field theories, including the $\phi^4$, are known to develop a non-trivial vacuum expectation value at critical coupling, if zero modes are taken into consideration.
The effect of their inclusion may result in the improved convergence of numerical results~\cite{PhysRevD.79.096012} and changes in the value of critical coupling~\cite{Burkardt:2016ffk,Chabysheva:2021yse,Chabysheva:2022duu}.}
Other advantages of the LF formulation include separation of internal and center-of-mass degrees of freedom; form-invariance of the Hamiltonian under Lorentz transformations; linearity of equations of motion, leading to a smaller number of independent field components; simple form of observables, balanced treatment of gauge and matter fields~\cite{BRODSKY1998299}.

\item In the treatment of a relativistic QFT, one typically starts with a \emph{canonical}  theory describing point-like interactions and operating with infinite ranges of momenta.
In order to obtain a numerically sensible non-perturbative \emph{effective} theory, one has to first \emph{regulate} the divergent interactions in the canonical theory and/or impose cutoffs on the Hilbert space dimension.
The relation between the values of coupling constants and observables in canonical and effective theories is governed by the renormalization group flow~\cite{wilson1975renormalization,wilson1971renormalization1,wilson1971renormalization2,wilson1972critical,glazek1993renormalization,glazek1994perturbative,wilson1994nonperturbative}.

Renormalization in the LF formalism can follow one of several approaches.
The original approach to renormalizing LF QFTs amounted to using the Pauli-Villars
regularization scheme~\cite{brodsky2003mass,hiller2003nonperturbative,brodsky2006two,chabysheva2009restoration,chabysheva2010nonperturbative,chabysheva2010nonperturbative,chabysheva2010nonperturbative,hiller2010pauli,chabysheva2011first,chabysheva2012light,chabysheva2015brst,hiller2016pauli,hiller2016nonperturbative,chabysheva2022nonperturbative}, in which the canonical Hamiltonian is regulated by introducing additional quantum fields, some of which have negative norm.
As a consequence of this, the canonical Hamiltonian turns into a non-Hermitian operator whose spectrum contains a number of unphysical states.
In the sector-dependent renormalization approach~\cite{perry1990light,perry1991renormalization,mathiot2010field,hiller1998nonperturbative,karmanov2008systematic,li2015ab,chabysheva2017light}, the canonical Hamiltonian operator is equipped with counterterms whose value depends on the number of particles in a Fock state the Hamiltonian acts upon.
Both Hamiltonian operator non-Hermiticity and sector dependence of its coefficients inevitably pose difficulties for quantum simulation.

A more systematic approach to renormalization, which is most suitable from the quantum simulation perspective, is pursued within RGPEP~\cite{glazek1993renormalization,glazek1994perturbative,wilson1994nonperturbative,glazek2012renormalization,glazek2012perturbative,glazek2013fermion,glazek2017renormalized,Serafin:2019vuk,glazek2021elementary,serafin2023dynamics} mentioned above in the context of the GWW flow.
In this case, the effective Hamiltonian operator is Hermitian and its action is defined in Fock sectors with arbitrary number of particles and without adding to the physical degrees of freedom.

\item In practice, constructing the operator $\OOPAd_{\Lambda,\,\bm{p}}$ will require switching between various reference frames.
In the non-relativistic setting, the wave functions of individual bound states are typically found in their respective center-of-mass frames, and then boosted into the center-of-mass frame of the combined system.
Similarly, the LF formulation of QFT allows one to find the wave functions of bound states using the so-called \emph{intrinsic coordinates}.
In~\Cref{sec:lfmanybody} we discuss construction of composite state wave functions in the LF dynamics.

\item \label{item:noncomm} While for operators $\OOPAd_{\Lambda_1,\,\bm{p}_1}$ and $\OOPAd_{\Lambda_2,\,\bm{p}_2}$, acting on disjoint sets of modes, the (anti-)symmetry properties of the state~\eqref{eq:manybint} would be satisfied automatically, in general this would not be the case.
The sensitivity of~\cref{eq:manybint} to the order of operators is the consequence of finiteness of momentum cutoffs in the theory, as the effective particle operators in RGPEP obey the same commutation relations as the original ones~\cite{glazek2011reinterpretation}.
This issue has to be addressed in the future work.

\item In~\Cref{sec:dlcq,sec:scatter_prob} we illustrate our ideas within the framework of Discretized Light-Cone Quantization (DLCQ)~\cite{pauli1,pauli2,phi4in2d,Brodsky:1997de}.
In particular, we shall demonstrate how our technique can be used to obtain ``exact'' results at exponential cost with a classical simulation.
Yet it is most naturally implemented by means of a quantum simulation, which we discuss in~\Cref{sec:scat_quant}.

\end{enumerate}

\section{Discretized Light-Cone Quantization\label{sec:dlcq}}

In this Section, we apply the construction introduced in~\Cref{sec:main_idea} to $\phi^4$ theory in $1+1\mathrm{D}$, formulated within the DLCQ~\cite{pauli1,pauli2,phi4in2d,BRODSKY1998299,Vary:2021cbh} framework.
Our primary objective here is to consider a scattering scenario in which the wavefunction of the combined system does not factorize, and operators creating composite particles act upon overlapping sets of modes.
We shall, therefore, ignore the issue of boosting the LF wavefunctions of composite particles (in the notations of~\Cref{sec:lfmanybody}, the wavefunctions discussed below correspond to $\sket{{\Psi}^{(l)} ({x}_i, {\bm{k}}_{\perp i})}$).
We adopt~\cref{eq:creatop} for defining operators which create particles of the interacting theory from vacuum, and use the exact unitary coupled cluster procedure, \cref{eq:eucc}, for finding the wave operator $\Wop$.

DLCQ is a discretized gauge-fixed ($A^+=0$, if gauge fields are present) Hamiltonian formulation of QFT, in which one quantizes the theory in a box, using the light-cone coordinates ${x^\pm=ct\pm x}$.
The  evolution of the system along the light-cone time $x^+$ is governed by operator $\mathcal{P}^-$, which in $1+1\mathrm{D}$ is related to the mass-squared operator as ${\mathcal{M}^2=\mathcal{P}^+\mathcal{P}^-}$, where $\mathcal{P}^+$ is the operator of LF momentum.
After rescalng the operators as ${\mathcal{P}^+=(2\pi/L)\mathcal{K}}$ and ${\mathcal{P}^-=(2\pi/L)^{-1}\mathcal{H}}$, where $L$ is the box size, the mass-squared operator can be written as ${\mathcal{M}^2=\mathcal{K}\mathcal{H}}$.
The operator of dimensionless discretized LF momentum $\mathcal{K}$ is termed \emph{harmonic resolution}, while $\mathcal{H}$ is typically referred to simply as the Hamiltonian, despite having the dimension of squared mass~\cite{pauli1,Brodsky:1997de}.

The normal-ordered Hamiltonian of the DLCQ $\phi^4$ model in ${1+1\mathrm{D}}$ with periodic boundary condition has the form~\cite{phi4in2d} (see also~\Cref{app:phi4ham}):
\begin{alignat}{9}
\label{eq:phi4ham}
    &\begin{alignedat}{9}
    \mathcal{H}_{\text{full}} &= \mathcal{H}_{\text{free}} + \mathcal{H}^{I} \, ,\\
    \mathcal{H}_{\text{free}} &= \mass^2 \sum_{\Moden=1}^{\infty}\frac{1}{\Moden}\OPad_\Moden \OPa_\Moden
    \, ,\\
    \mathcal{H}^{I}&=\frac{1}{4}\frac{\lambda}{4\pi}\sum_{\Modek \Model  \Modem\Moden=1}^{\infty}\frac{\OPad_\Modek \OPad_\Model \OPa_\Modem \OPa_\Moden}{\sqrt{\Modek\Model\Modem\Moden}}&&\delta_{\Modem+\Moden,\Modek+\Model}\\
&+\frac{1}{6}\frac{\lambda}{4\pi}\sum_{\Modek\Model\Modem\Moden=1}^{\infty}\frac{\OPad_\Modek \OPa_\Model \OPa_\Modem \OPa_\Moden}{\sqrt{\Modek\Model\Modem\Moden}}&&\delta_{\Modek,\Modem+\Moden+\Model}\\
&+\frac{1}{6}\frac{\lambda}{4\pi}\sum_{\Modek\Model\Modem\Moden=1}^{\infty}\frac{\OPad_\Moden \OPad_\Modem \OPad_\Model \OPa_\Modek}{\sqrt{\Modek\Model\Modem\Moden}}&&\delta_{\Modek,\Modem+\Moden+\Model} \, .
    \end{alignedat}
    \\
    &\hspace{0.4cm}\Kop = \sum \limits_{\Moden=1}^\infty \Moden (\OPad_\Moden \OPa_\Moden ) \, .
\end{alignat}

The Canonical Sets of Commuting Observables (CSCOs) of the free and interacting theories consist of operators~$\{\Hopfree,\,\Kop,\,\mathcal{N}\}$ and~$\{\Hopfull,\,\Kop,\,\mathcal{Z}\}$, correspondingly, where ${\mathcal{N}=\sum_{\Moden=1}^\infty \OPad_\Moden \OPa_\Moden}$ is the number operator, while the operator ${\mathcal{Z}=\mathcal{N}(\!\!\!\!\mod 2)}$ marks the sectors of odd and even number of particles.
Neither $\mathcal{N}$ nor $\mathcal{Z}$ will play a role in the following discussion.

The Hamiltonian~\eqref{eq:phi4ham} is solved in the basis of Fock states of the form
\begin{equation}
\begin{gathered}
    \Fock = \sket{n_1^{\occ_1},\,n_2^{\occ_2},\,n_3^{\occ_3},\ldots} \,,\\
    n_j,\,\occ_j=1,2,3,\ldots
\end{gathered}
\end{equation}
where $n_j$ are the momentum quantum numbers and $\occ_j$ are the occupancies.

The Fock space splits into blocks of fixed harmonic resolution ${\Kop=K}$ and fixed even or odd particle number~\cite{phi4in2d}.
These blocks have finite dimension owing to the fact that all the states within each block contain particles whose momenta are positive integers summing up to $K$:
\begin{equation}
    \Fock[][K]:\quad
    \quad\sum_j n_j \occ_j = K \, .
\end{equation}
Using more general notation (see~\Cref{sec:definitions}), one can say that each such block is obtained by restricting the infinite-dimensional Fock space to a subspace comprised of vectors with momentum modes in $\Modeset$, with the cutoff on the maximum number of excitations in each momentum mode $\Moden$ given by $\occmax(\Moden)$, and with the maximum number of modes in a state being~$\maxmodes$:
\begin{subequations}
\begin{gather}
    \Modeset: \quad \{1,2,\ldots,K\} \, ,\\
    \occmax(\Moden) = \lfloor K/\Moden\rfloor \, ,\\
    \maxmodes = K \, .
\end{gather}
\end{subequations}

The number of Fock states at a fixed value of~$K$ is exactly equal to the number of integer partitions~$p(K)$, which grows as~${p(K)=\Theta(\exp(\sqrt{K}))}$~\cite{andrews1998theory,Kreshchuk:2020dla}.

The versions of~$\Hopfree[K]$ and~$\Hopfull[K]$, whose action is restricted to momentum modes up to $K$, are obtained by choosing $K$ as the cutoff for the sums in~\cref{eq:phi4ham}.
Within a sector of fixed $K$, diagonalizing ${\mathcal{M}^2=\Kop\mathcal{H}_{\mathrm{full}}}$ is equivalent to diagonalizing $\mathcal{H}_{\mathrm{full}}$.
Determining the spectrum of $\mathcal{M}^2$ at higher values of $K$ may be interpreted as studying the system at a higher resolution, which explains the notion of \emph{harmonic resolution}~\cite{pauli1,pauli2}.

In the DLCQ treatment of ${1+1\mathrm{D}}$ models, one typically does not renormalize the coupling constant~\cite{pauli1,pauli2,phi4in2d}, while
mass renormalization is performed by adjusting the bare mass $\mass$ in~\cref{eq:phi4ham} so that for each value of $K$ the lowest eigenvalue of $\mathcal{M}^2$ remains unchanged~\cite{phi4in2d}.
This, however, implies that the value of bare mass depends on $K$, which is not compatible with our approach.
For the same reason, neither is the sector-dependent renormalization discussed in~\Cref{sec:main_idea}).
A better option would be to first renormalize the original continuous theory using RGPEP, and then to solve it using DLCQ~\cite{glazek1993renormalization,glazek1994perturbative,wilson1994nonperturbative,glazek2012renormalization,glazek2012perturbative,glazek2013fermion,Serafin:2019vuk,glazek2021elementary}.
In this paper we assume the values of both $\mass$ and $\lambda$ to be fixed.

Note that in order to reproduce the action of~$\Hopfull$ on a state $\Fock[][K]$, one has to include all the terms acting on momentum modes up to $K$:
\begin{subequations}
\label{eq:hhkk}
\begin{alignat}{9}
&\Hopfull[K^\prime] &&\Fock[][K] &&= &&\Hopfull \Fock[][K] \, &&,
\quad K^\prime \geq K\,, \\
&\Hopfull[K^\prime] &&\Fock[][K] &&\neq \ &&\Hopfull \Fock[][K] \, &&,
\quad K^\prime < K\, .
\end{alignat}
\end{subequations}

We denote the matrices of~$\Hopfree$ and~$\Hopfull$ (or, equivalently, of~$\Hopfree[K]$ and~$\Hopfull[K]$) in the basis of~$\Fock[][K]$
by~$\Hmatfree[K]$ and~$\Hmatfull[K]$.
At a fixed value of harmonic resolution, the matrix~$\Hmatfull[K]$ can be further block-diagonalized, owing to the fact the self-interaction in~\cref{eq:phi4ham} either preserves the particle number in a Fock state, or changes it by two.
Thereby, the Hilbert space splits into the \textit{even} and \textit{odd} sectors, in which the Fock states contain either even or odd number of particles, correspondingly~\cite{phi4in2d}.

We adopt the standard normalization of Fock states~\cite{phi4in2d,pauli1,pauli2}
\begin{equation}
    \label{eq:focknorm}
    \sbraket{\text{vac}}{\text{vac}}
    =
    \sbraket{\FF}{\FF}
    = 1 \, .
\end{equation}
Similar to non-relativistic many-body theory, the LF vacuum state is simply a state without any excitations, and it is the only state with total LF momentum zero~\cite{pauli1,Brodsky:1997de}.

Let operator~$\OPtAd[K][n]$ create the $n$th (in the order of increasing $\Hopfree$ eigenvalue) Fock state $\Fock[][K][n]$.
It is a monomial in single-mode creation operators $\OPad_\Moden$, with the constant fixed by~\cref{eq:focknorm}.
For the ground states $\Fock[][K][0]$ in sectors of even and odd numbers of particles, operators creating those from vacuum will be denoted by $\OOPad[even]_{[K,0]}$ and $\OOPad[odd]_{[K,0]}\equiv\OOPad_{[K,0]}$\footnote{
The convention $\OOPad[odd]_{[K,0]}\equiv\OOPad_{[K,0]}$ is justified by the fact that for each $K$, the ground state energy in the odd sector of the free Hamiltonian is always lower than the ground state in the even sector, see~\cref{eq:phi4ham}.
}.
The ground states as well as the corresponding creation operators expressed in terms of $\OPad_\Moden$ are shown in~\Cref{tab:GSs}.

\begin{table*}
\centering
\def\arraystretch{1.4}
\begin{tabular}{|l|cc|}
\hline
\multirow{2}{*}{\makecell{Number of particles\\in Fock states}} & \multicolumn{2}{c|}{\makecell{Ground states of $\Hopfree$ at $\Kop=K$, and operators creating those from vacuum}}\\ \cline{2-3}
& \multicolumn{1}{c|}{Even $K$} & Odd $K$
\\ \hline
\multirow{2}{*}{Even} & \multicolumn{1}{c|}{$\sket{(\frac{K}{2})^2} = \OOPad[even]_{[K,0]} \vac$}  & $ \sket{\frac{K-1}{2},\frac{K+1}{2}}= \OOPad[even]_{[K,0]} \vac $ \\ \cline{2-3}
& \multicolumn{1}{c|}{$\OOPad[even]_{[K,0]} = 1/\sqrt{2!} \, (\OPad_{K/2})^2$} & $ \OOPad[even]_{[K,0]}=\OPad_{(K+1)/2}\OPad_{(K-1)/2}$ \\\hline
\multirow{2}{*}{Odd} & \multicolumn{2}{c|}{${\sket{K}} = \OOPad[odd]_{[K,0]} \vac$} \\ \cline{2-3}
& \multicolumn{2}{c|}{${\OOPad_{[K,0]}\equiv\OOPad[odd]_{[K,0]}=\OPad_K}$}  \\\hline
\end{tabular}
\caption{Top cells: ground states in the sectors of even and odd numbers of particles of the free $\phi^4_{1+1}$ theory in the DLCQ formulation, for even and odd values of harmonic resolution $K$.
Bottom cells: operators creating these ground states from vacuum, expressed in terms of single-mode creation operators.}
\label{tab:GSs}
\end{table*}
\begin{sloppypar}

At a fixed value of $K$, the interacting eigenstates $\FFock[][K]$ are obtained by diagonalizing the matrix~$H_{\text{full}, K}$ in the basis of $\Fock[][K]$.
The two orthonormal bases~$\Fock[][K]$ and~$\FFock[][K]$ are related by a unitary matrix~$\W[K]$, comprised of the eigenvectors of~$H_{\text{full},K}$, written in the basis of Fock states~$\FFock[][K]$ (\emph{modal matrix}):\footnote{Unless stated otherwise, the sets~$\bigl\{\Fock[][K]\bigr\}$ and~$\bigl\{\FFock[][K]\bigr\}$ are assumed to contain states from both even and odd particle number sectors.}

\begin{equation}
    \label{eq:basis_rot}
    \begin{alignedat}{8}
        \FFock[][K][n] &= \sum_m(\W[K])_{n}{}^{\!m} \Fock[][K][m] \, ,
    \end{alignedat}
\end{equation}
where $\FFock[][K][n]$ stands for the $n$th state.
We assume the ordering of states~$\FFock[][K][n]$ to match that one given by the adiabatic interaction turn-on, so that $\FFock[][K][n]$ is the adiabatic continuation of $\Fock[][K][n]$.
For a weakly-coupled theory, the free and interacting eigenstates could be matched by sorting both $\Fock[][K][n]$ and $\FFock[][K][n]$ in the order of growing $\Hopfree[K]$ and $\Hopfull[K]$ eigenvalues.
For a strongly-coupled theory, one could match the free and interacting basis states using an approximate implementation of the adiabatic turn-on.
\end{sloppypar}

While~\eqref{eq:basis_rot} should be understood as a \emph{matrix equation}, which holds within a block of fixed~$K$, we would now like to find a wave operator operator~$\Wop[K]$, whose action on~$\bigl\{\Fock[][K]\bigr\}$ is given by~$\W[K]$.

By writing~$\Wop[K]$ in terms of single-mode creation and annihilation operators, we shall \emph{extend} the action of~$\W[K]$ to Fock states~$\bigl\{\Fock\bigr\}$ from sectors of arbitrary harmonic resolution.
While, strictly speaking, there is no unique way of defining such an extension, below we explicitly construct operator~$\Wop[K]$, acting as wave operator for sectors of Hilbert space or harmonic resolution up to~$K$.

We begin by defining a Hermitian operator ${\Vop[K] = \Vopd[K]}$ of the form
\begin{equation}
        \label{eq:WV}
        \Wop[K] = \me^{-\iu \Vop[K]} \, .
\end{equation}
We represent $\Vop[K]$ as a polynomial in single-mode creation operators of the free theory, $\Vop[K] = \poly(\OPad_\Moden,\OPa_\Moden)$.
Then, we postulate the following defining properties of $\Wop[K]$:
\begin{enumerate}
    \item The operator $\Wop[K]$ is unitary:
    \begin{equation}
        \label{eq:unit}
        \Wopd[K]\Wop[K]= \unit \, .
    \end{equation}

    \item The action of $\Wop[K]$ on~$\bigl\{\Fock[][K][n]\bigr\}$ is given by~\eqref{eq:basis_rot}:
    \begin{equation}
    \label{eq:act}
        \Wop[K] \Fock[][K][n] =
        \W[K] \Fock[][K][n] = \FFock[][K][n] \, ,
    \end{equation}
    i.e. $\W[K]$ is the matrix of the $\Wop[K]$ operator in the basis of~$\bigl\{\Fock[][K][n]\bigr\}$.

    \item For the operator $\Vop[K]$, the following holds:
    \begin{equation}
        \label{eq:vset}
        \Vop[K] \supseteq \Vop[K-1] \supseteq \ldots \supseteq \Vop[2] \supseteq \Vop[1] \, ,
    \end{equation}
    where the notation $\Vop[j] \supseteq \Vop[j-1]$ for polynomials $\Vop[j] $ and $\Vop[j-1]$ means that $\Vop[j]$ contains all the terms from $\Vop[j-1]$.
    With \cref{eq:vset} we require that~\eqref{eq:act} holds for any $K^\prime \leq K$:
    \begin{equation}
    \label{eq:actprime}
        \Wop[K] \Fock[][K'][n] =
        \W[K'] \Fock[][K'][n] \quad \text{for } K'\leq K
    \end{equation}

    \item Operator $\Vop[K]$ contains a minimal number of terms required to satisfy $\eqref{eq:act}$.
\end{enumerate}
With properties 1 and 2 we ensure that $\Wop[K]$ is the wave operator in sector of momentum $K$.
With properties 3 and 4 we additionally require that is $\Vop[K]$ has the simplest possible form, such that $\Wop[K]$ is also the wave operator in sectors of momenta below $K$.
In other words, we require the matrix of~$\Wop[K]$ in the basis of all the Fock states~$\bigl\{\Fock[][\leq K]\bigr\}$ of momenta up to $K$ to be equal to~$\diag\{\W[1],\ldots,\W[K]\}$.
As follows from the definition above, $\Wop[K]$ can be thought of as an ``exact'' wave operator for any ${K^\prime \leq K}$ and as its ``reduced'' version for ${K^\prime > K}$.
It is interesting to note that results for $\lambda \phi^4$ in $1+1\mathrm{D}$ from multiple values of $K$ were employed simultaneously to calculate form factors and to search for kink condensation in~\cite{chakrabarti2005transition}.

In order to explicitly construct~$\Wop[K]$,
we seek~$\Vop[K]$ in the form of
\begin{equation}
\label{eq:uccfull}
\begin{multlined}
    \Vop[K] = \!\sum_{r,s=1}^K
    \sum_{\substack{
    \Modei_1,\Modei_2\ldots,\Modei_r
    \\
    \Modej_1,\Modej_2\ldots,\Modej_s
    }}
\!\!\!\theta_{\Modei_1,\ldots,\Modei_r,\Modej_1,\ldots,\Modej_s}
    \OPad_{\Modei_1}\ldots\OPad_{\Modei_r}
    \OPa_{\Modej_1}\ldots\OPa_{\Modej_s} \, ,
\end{multlined}
\end{equation}
whose normal-ordered form automatically guarantees that ${\Wop[K]\vac=\vac}$, and which has enough free parameters to ensure that the matrix of~$\Wop[K]$ coincides with~$\diag\{\W[1],\ldots,\W[K]\}$ in the basis of~$\bigl\{\Fock[][\leq K]\bigr\}$.
It is assumed that $\Vop[K]$ commutes with the symmetry operators of the system (such as $\Kop$), which ensures that so does $\Wop[K]$.
While eq.~\eqref{eq:uccfull} contains, in principle, all the monomials whose matrix elements are non-zero for \emph{some} pair of Fock states, the number of terms in it is excessive.
Indeed, physically, we expect $\Wop(\Modeset,\occmax,\maxmodes)$ to depend on $\occmax$ and $\maxmodes$ only due to the cutoff artifacts~--- which are absent in ${1+1\mathrm{D}}$ LF QFTs (meaning that the Hilbert space of the Hamiltonian~\eqref{eq:phi4ham} is finite for a fixed value of $K$).
We describe the algorithm for finding the coefficients in~\cref{eq:uccfull} in~\Cref{app:ucc}.

The states $\FFock[][K][n]$ can be created from vacuum upon the application of operators ${\OPtAd[K][n]}$, which we define~as
\begin{equation}
        \label{eq:OPtad}
        \OPtAd[K][n] = \bigl(\Wop[K]\bigr)\bigl(\OPtad[K][n]\bigr) \bigl(\Wop[K]\bigr)^\dagger \, .
\end{equation}
Indeed, due to the uniqueness of the LF vacuum, one can write:
\begin{equation}
\label{eq:OPtadvac}
\begin{alignedat}{8}
    \OPtAd[K][n] \vac &=
    \bigl(\Wop[K]\bigr)\bigl(\OPtad[K][n]\bigr) \bigl(\Wop[K]\bigr)^\dagger \vac \\
    &=
    \bigl(\Wop[K]\bigr)\bigl(\OPtad[K][n]\bigr) \vac \\&=
    \bigl(\Wop[K]\bigr) \Fock[][K][n]  = \FFock[][K][n] \, .
\end{alignedat}
\end{equation}

\begin{sloppypar}
Generally speaking, in order to simulate non-trivial time evolution within a Hamiltonian formulation of QFT, one has to either initialize the system in an eigenstate and then switch on an external field~\cite{PhysRevD.88.065014,PhysRevD.101.076016,barata2023quantum} or prepare an initial state, which is a non-stationary superposition of eigenstates~\cite{PhysRevD.88.065014,PhysRevC.97.064620,PhysRevA.104.012611,Lei:2022nsk}.
We follow the latter approach by introducing the notation
\begin{equation}
\label{eq:consec2}
\begin{alignedat}{8}
    &\hspace{-0.5cm}
    \mathrlap{\bket{[K_1,n_1]^{\occint_1},[K_2,n_2]^{\occint_2},\ldots}}
    \\
    \equiv& \
    \Dconst_{\{[K_1,n_1]^{\occint_1},[K_2,n_2]^{\occint_2},\ldots\}}
    \\
    &\times
    \bigl(\OPtAd[K_1][n_1]\bigr)^{\occint_1}
    \bigl(\OPtAd[K_2][n_2]\bigr)^{\occint_2} \ldots\vac \, ,
\end{alignedat}
\end{equation}
for a state in the full theory with ``$\occint_1$ particles of momentum $K_1$ in the $n_1$th excited state, $\occint_2$ particles of momentum $K_2$ in the $n_2$th excited state, etc.'', carrying the total momentum ${K_{\text{tot}}= \sum_j K_j \occint_j }$.
Unlike the analogous state
$\sket{n_1^{\occ_1},n_2^{\occ_2},\ldots}$
of the free system, which \emph{is} an eigenstate of $H_{\text{free},K_{\text{tot}}}$, the state~\eqref{eq:consec2} is generally \emph{not} an eigenstate of either $H_{\text{free},K_{\text{tot}}}$ or $H_{\text{full},K_{\text{tot}}}$.\footnote{
Note that, while the eigenstates of $H_{\text{full},K_{\text{tot}}}$ are generally the superpositions of Fock states containing modes of LF momentum up to $K_{\text{tot}}$, in~\eqref{eq:consec2} only modes of LF momentum up to $\max\{K_j\}$ are included. However, not containing single-particle momenta above $\max\{K_j\}$ is \emph{not} by itself a reason for $\bket{[K_1,n_1]^{\occint_1},[K_2,n_2]^{\occint_2},\ldots}$ to not be an eigenstate of $H_{\text{full},K_{\text{tot}}}$.
Indeed, while $\sket{n_1^{\occ_1},n_2^{\occ_2},\ldots}$ similarly does not carry LF momenta higher than $\max\{K_j\}$, it \emph{is} an eigenstate of $H_{\text{free},K_{\text{tot}}}$.
What matters is that $\Wop[K_{\text{tot}}]$ has not been used in the definition~\eqref{eq:consec2}.
}
\end{sloppypar}

A state containing two particles of momenta $K_1$ and $K_2$, each in the ground state of the corresponding Hamiltonian, has the form of
\begin{equation}
    \label{eq:K1K2}
    \begin{alignedat}{8}
    &
    \bket{i} =
    \bket{[K_1,0],[K_2,0]}\\
    &=
    \Dconst_{\{[K_1,0],[K_2,0]\}}
    \bigl(\OPtAd[K_1,0]\bigr) \bigl(\OPtAd[K_2,0]\bigr) \vac
    \\
    &=
    \Dconst_{\{[K_1,0],[K_2,0]\}}
    \Wop[K_1] \OPad_{K_1} \Wopd[K_1]
    \Wop[K_2] \OPad_{K_2} \Wopd[K_2] \vac
    \\
    &=
    \Dconst_{\{[K_1,0],[K_2,0]\}}
    \Wop[K_1] \OPad_{K_1} \Wopd[K_1]
    \Wop[K_2] \OPad_{K_2} \vac \, .
    \end{alignedat}
\end{equation}

A particularly simple case of~\eqref{eq:K1K2} is obtained by setting $K_1=K_2=K$:
\begin{equation}
    \label{eq:2K}
\begin{alignedat}{8}
    \nfull[K][0][2] &= \Dconst_{\{[K,0]^2\}} \bigl(\OPtAd[K,0]\bigr)^2 \vac \\
    &= \Dconst_{\{[K,0]^2\}} \Wop[K] \bigl(\OPad_{K}\bigr)^2 \vac
    \, .
\end{alignedat}
\end{equation}
The so-defined state is not an eigenstate of~$\Hopfull$, as $\Wop[K]$ would only produce an eigenstate of~$\Hopfull$ \emph{when acting on states of momentum up to $K$}.
The state in~\cref{eq:2K} should not be confused with the even sector ground state of $H_{\text{full},2K}$:
\begin{equation}
    \label{eq:2Knot}
\begin{multlined}
    \nfull[K][0][2] \neq
    \nfull[2K][0]
    \\=\Dconst_{\{[2K,0]\}} \bigl(\Wop[2K]\bigr) \bigl(\OPad_{K}\bigr)^2 \vac \, .
\end{multlined}
\end{equation}

For elastic and inelastic ${2\to2}$ scattering processes, we define the final states to be
\begin{subequations}
\label{eq:finalscat}
\begin{alignat}{7}
    \label{eq:elastic}
    \sket{f_{\mathrm{elastic}}} &= \bket{[K_1^\prime,0],[K_2^\prime,0]}
    \, ,\\
    \label{eq:inelastic}
    \sket{f_{\mathrm{inelastic}}} &= \bket{[K_1^\prime,n_1],[K_2^\prime,n_2]} \, ,
\end{alignat}
\end{subequations}
where ${K_1 + K_2 = K_1^\prime+K_1^\prime}$ and ${n_1+n_2 > 0}$ is assumed in the second line.

Upon reviewing the DLCQ formulation of $\phi^4$ theory in $1+1\mathrm{D}$, we introduced the operator $\Wop[K]$ relating the bases of the free and interacting theories.
Using this operator, in~\eqref{eq:OPtad} we defined operators ${\OPtAd[K][n]}$ creating eigenstates of the interacting theory from vacuum.
We then used these operators to define in~\eqref{eq:consec2} the multi-particle states in the interacting theory.
In the following Section, time evolution of such states will be studied.

\section{Classical Simulation of Time Evolution\label{sec:scatter_prob}}

\begin{table*}
\begin{tabular}{|l|l|l|l|l|}
\hline
\makecell{Harmonic\\resolution, $K$} & Fock states & Hamiltonian matrix & Modal matrix, $W_K$ & $\VV[K]=\iu\ln\bigl(\W[K]\bigr)$ \\ \hline
1, odd sector &
$\sket{1}$ &
$\begin{pmatrix} 0.5 \end{pmatrix}$ &
$\begin{pmatrix} 1 \end{pmatrix}$ &
$\begin{pmatrix} 0 \end{pmatrix}$
\\ \hline
2, odd sector &
$\sket{2}$ &
$\begin{pmatrix} 1/2 \end{pmatrix}$ &
$\begin{pmatrix} 1 \end{pmatrix}$ &
$\begin{pmatrix} 0 \end{pmatrix}$
\\ \hline
2, even sector&
$\sket{1^2}$  &
$\begin{pmatrix} 2+15/(4\pi) \end{pmatrix}$ &
$\begin{pmatrix} 1 \end{pmatrix}$ &
$\begin{pmatrix} 0 \end{pmatrix}$
\\ \hline
$3$, odd sector&
$\begin{matrix}
\sket{3} \\ \sket{1^3}
\end{matrix}$ &
$\begin{pmatrix}
1/3 & 5\sqrt{2}/(4\pi) \\
5\sqrt{2}/(4\pi) & 3+45/(4\pi)
\end{pmatrix}$
&
$\begin{pmatrix}
0.996 & 0.089 \\
-0.089 & 0.996
\end{pmatrix}$ &
$\begin{pmatrix}
0 & 0.0089 \\
-0.0089 & 0
\end{pmatrix}$
\\ \hline
3, even sector &
$\sket{1,2}$&
$\begin{pmatrix} 3/2+15/(8\pi) \end{pmatrix}$  &
$\begin{pmatrix} 1 \end{pmatrix}$ &
$\begin{pmatrix} 0 \end{pmatrix}$
\\ \hline
\end{tabular}
\caption{
Hamiltonians ($\Hmatfull[K]$) and modal matrices ($\W[K]$) for $K\leq3$.
The parameters of the UCC operator
${\Vop[K]=\iu\ln{\bigl(\Wop[K]\bigr)}}$ in eq.~\eqref{eq:uccfull}
are found by equating the matrix elements of ${\Vop[K^\prime]}$ in the basis of Fock states to those of the $\VV[K^\prime]=\iu\ln\bigl(\W[K^\prime]\bigr)$ for all $K^\prime\leq K$.
Note that the diagonal entries of $\W[K]$ are set positive in order for the matrix logarithm to be defined.
This can be always achieved by adjusting the phase of the eigenvector.
}
\label{tab:w}
\end{table*}

\begin{sloppypar}
In this Section, we calculate the time-dependent transition probability $\bigl|\sand{f}{\me^{-\iu \Hopfull t}}{i}\bigr|^2$ for an elastic scattering process, in which the initial state is defined as in~\cref{eq:2K} with ${K_1=K_2=3}$ and the final states are defined as in~\cref{eq:K1K2} with ${K_1+K_2=6}$.
We set the constants in the Hamiltonian~\eqref{eq:phi4ham} to be ${\mass = 1}$ and ${\lambda=30}$.
\end{sloppypar}

Following~\cref{eq:2K}, we prepare the initial state ${\bket{i} = \bket{[3,0]^2}}$ with the aid of the $\Wop[3]$ operator, which is constructed  by means of the unitary coupled cluster~\cite{shen2017quantum,Romero_2018,filip2020stochastic} procedure.
First, we list all the modal matrices $\W[K]$ for $K\leq3$, see~\Cref{tab:w}, and find such an operator $\Vop[3]$ that for any $K\leq3$ its matrix in the basis of Fock states $\bigl\{\Fock[][K]\bigr\}$ is $\VV[K]\equiv\iu\ln\bigl(\W[K]\bigr)$ (see~\Cref{app:ucc} for details):
\begin{equation}
    \Vop[3] = 0.0364 \iu \bigl(
  (\OPad_1)^3 \OPa_3 - \OPad_3 (\OPa_1)^3 \bigr) \ .
\end{equation}

\begin{sloppypar}
Next, we define ${\Wop[3] = \me^{-\iu \Vop[3]}}$  and $\OPtAd[3,0]=\Wop[3]\OPad_3\Wopd[3]$, which relate the free and interacting ground states at $K=3$ as follows:
\begin{alignat}{9}
    &\OPtad[3,0]\vac = \OPad_3\vac = \sket{3} \ ,\\
&
\label{eq:a30}
\begin{alignedat}{8}
\OPtAd[3,0]
    &=
    \Wop[3] \OPtad[3,0] \Wopd[3]
    = \ &&
    \me^{0.0364 \iu (
    (\OPad_1)^3 \OPa_3 - \OPa_3 (\OPad_1)^3 )}\OPad_3
    \\
    &&&\mathllap{\times}
    \me^{- 0.0364 \iu (
    \OPad_3 (\OPa_1)^3 - (\OPa_1)^3 \OPad_3 )} \\
\end{alignedat}
\\
\label{eq:30}
&\begin{alignedat}{8}
    \bket{[3,0]} &= \OPtAd[3,0] \vac
    \\
    &=
    \mathrlap{\bigl( 0.996 \OPad_3 - (\OPad_1)^3/(\sqrt{6})\bigr)\vac}
    \\
    &= 0.996 \sket{3} - 0.089 \sket{1^3} \, .
\end{alignedat}
\end{alignat}

Note that if, instead of $\OPtAd[3,0]$ defined in~\cref{eq:a30}, had we used the polynomial
$\OPtPd[{[3,0]}]$ to create particles in the interacting theory, the latter, according to~\cref{eq:30}, would have acquired the form of
\begin{equation}
    \label{eq:pK}
    \OPtPd[{[3,0]}] = 0.996 \OPad_3 - (\OPad_1)^3/(\sqrt{6}) \,.
\end{equation}
While $\OPtPd[{[3,0]}]$
and $\OPtAd[3,0]$
act identically on $\vac$,
\begin{equation}
    \OPtPd[{[3,0]}] \vac = \OPtAd[3,0] \vac \,,
\end{equation}
they, generally, act differently on arbitrary Fock states, see~\Cref{tab:states}.
Consider, as an example, the state $\bket{[3,0]^2}$, containing two particles of momentum $3$.
Whether it is defined as $\bigl( \OPtAd[3,0] \bigr)^2\vac$ or $\bigl( \OPtPd[{[4,0]}] \bigr)^2\vac$, it will belong to the space ${\Span\hspace{-0.5pt}\bigl\{\sket{3^2},\sket{1^3,3},\sket{1^6}\bigr\}}$.
However, the amplitudes in the two cases will not be equal.
The difference becomes more pronounced at $K=4$, where, due to the presence of terms such $\OPad_1\OPad_3\bigl(\OPa_2\bigr)^2$ in $\Vop[4]$, the state $\bigl( \OPtAd[4,0] \bigr)^2\vac$ belongs to a subspace spanned over a larger number of Fock vectors than $\bigl( \OPtPd[{[4,0]}] \bigr)^2\vac$:
\begin{subequations}
\begin{alignat}9
&\begin{alignedat}9
    \bigl( \OPtAd[4,0] \bigr)^2&\vac \subset \Span\hspace{-0.5pt}\bigl\{
    \sket{4^2},\sket{1^2,2,4},\sket{1^4,2},
    \\&\sket{1^2,3^2},\sket{1,2^2,3},\sket{2^4},\sket{1^5,3},\sket{1^8}
    \bigr\}\,,
\end{alignedat}
    \\
    &\bigl( \OPtPd[{[4,0]}] \bigr)^2\vac \subset \Span\hspace{-0.5pt}\bigl\{
    \sket{4^2},\sket{1^2,2,4},\sket{1^4,2}
    \bigr\}\,.
\end{alignat}
\end{subequations}

Similarly to $\Wop[3]$, we find the wave operators $\Wop[2]$, $\Wop[4]$, and $\Wop[5]$, and use those to define the initial and final states according to~\eqref{eq:K1K2}.
The amplitudes of the resulting states in the basis of ${K=6}$ Fock states are shown in~\Cref{tab:states}, as well as the states obtained with the aid of operators $\OPtPd[{[n,0]}]$.

Finding the exact $\Wop[K]$ is a costly procedure:  calculating the modal matrix requires diagonalizing the Hamiltonian matrix whose dimension is exponential in $K$, and the number of free parameters in $\Vop[K]$ grows exponentially with $K$ as well.
These parameters can be found via solving a linear system of equations, as described in~\Cref{app:ucc}.
\end{sloppypar}

\begin{table*}[]
\begin{tikzpicture}
\hspace{-0.3cm}
\node (table) {\def\arraystretch{1.7}
\begin{ruledtabular}
\begin{tabular}{l||l||cccccc||c}
& & $\bket{3^2}$ & $\sket{2,4}$ & $\sket{1,5}$ & $\sket{1^2,2^2}$ & $\sket{1^3,3}$ & $\sket{1^6}$ & $\spec H_{\text{full,}K=6,\text{even}}$ \\ \hline\hline
1. & $\bket{[6,0]} = \OPtAd[6,0]\vac$           & 0.85962 & \llap{-}0.50001  &  0.00766 &  0.05678 & \llap{-}0.08757 &  0.00946 & 0.60     \\ \hline
2. & $\bket{[6,1]} = \OPtAd[6,1]\vac$           & 0.4483  &  0.75733  & \llap{-}0.47253 & \llap{-}0.04662 &  0.00483 & \llap{-}0.00053  & 0.88  \\ \hline
3. & $\bket{[6,2]} = \OPtAd[6,2]\vac$           & 0.22456 &  0.40704  &  0.87382 & \llap{-}0.09627 & \llap{-}0.10448 &  0.01188  & 1.77   \\ \hline
4. & $\bket{[6,3]} = \OPtAd[6,3]\vac$           & 0.0797  & \llap{-}0.06919  &  0.03475 & \llap{-}0.62988 &  0.75888 & \llap{-}0.12248  & 8.31  \\ \hline
5. & $\bket{[6,4]} = \OPtAd[6,4]\vac$           & 0.05717 &  0.0773   &  0.10884 &  0.76713 &  0.6147  & \llap{-}0.11208 & 10.10   \\ \hline
6. & $\bket{[6,5]} = \OPtAd[6,5]\vac$           & 0.00569 &  0.00049  &  0.00584 &  0.00955 &  0.16624 &  0.986  & 24.33    \\ \hline \hline
7. & $\bket{[5,0],[1,0]} = \OPtAd[5,0] \OPtAd[1,0] \vac$     & 0.00002 & \llap{-}0.01073  &  0.98935 & \llap{-}0.13779 &  0.03858 & \llap{-}0.02455 &  \\ \hline
8. & $\bket{[4,0],[2,0]} = \OPtAd[4,0] \OPtAd[2,0] \vac$     &\llap{-}0.00473 &  0.99258  &  0.      & \llap{-}0.06986 & \llap{-}0.09874 &  0.01151 &  \\ \hline
9. & $\bket{[3,0]^2} = \bigl( \OPtAd[3,0] \bigr)^2 \vac$         & 0.99217 &  0.       &  0.      &  0.      & \llap{-}0.12238 &  0.02475 &  \\ \hline
10. & $\bket{[2,0],[4,0]} = \OPtAd[2,0] \OPtAd[4,0] \vac$     & 0.      &  0.98989  &  0.      & \llap{-}0.14182 &  0.      &  0.  &      \\ \hline
11. & $\bket{[1,0],[5,0]} = \OPtAd[1,0] \OPtAd[5,0] \vac$     & 0.      &  0.       &  0.98653 & \llap{-}0.13576 & \llap{-}0.09109 & \llap{-}0.00611 &  \\ \hline\hline
12. & $\OPtPd[{[1,0]}] \OPtPd[{[5,0]}] \vac$   & 0. &      0.   &    0.98653 & -0.13576 & -0.09109 & -0.00611 & \\ \hline
13. & $\OPtPd[{[2,0]}] \OPtPd[{[4,0]}] \vac$   & 0. &      0.98989 &  0.  &    -0.14182 & 0.  &     0. & \\ \hline
14. & $\bigl(\OPtPd[{[3,0]}]\bigr)^2 \vac$   & 0.9918 &  0. &       0. &      0. &      -0.12532 &  0.02504 & \\ \hline\hline
15. & $\bket{\Psi_{\rm eq}}$   & 0.40825 & 0.40825   &  0.40825 & 0.40825  & 0.40825  & 0.40825 & \\ \hline\hline
16. & $\operatorname{spec} H_{\text{free,}K=6,\text{even}}$ & 0.66666 & 0.75 & 1.2 & 3 & 3.33333 & 6 &
\end{tabular}
\end{ruledtabular}
};
\draw [red,ultra thick,rounded corners]
  ($(table.south west) !.2! (table.north west) + (5.57,-0.7)$)
  rectangle
  ($(1.12,1.05)$  );
\end{tikzpicture}
    \caption{Rows 1-6: amplitudes of the eigenstates of the $H_{\text{full,}K=6,\text{even}}$ Hamiltonian in the basis of ${K=6}$ Fock states, as well as the corresponding eigenenergies (in the rightmost column).
    Rows 7-11: amplitudes of the two-particle composite states ${\OPtAd[n,0]\OPtAd[K-n,0]\vac = \bket{[n,0],[K-n,0]}}$ in the basis of ${K=6}$ Fock states.
    Rows 12-14: amplitudes of the two-particle composite states ${\OPtPd[{[n,0]}] \OPtPd[{[K-n,0]}] \vac}$ in the basis of ${K=6}$ Fock states.
    Row 15: $\bket{\Psi_{\rm eq}}$, the equal weight superposition state of Fock states in the ${K=6}$ even sector.
    Row 16: The eigenvalues of the $H_{\text{free,}K=6,\text{even}}$ eigenstates.
    The composite states largely overlap  with the three lowest states of both free and interacting theories (highlighted with the box), which explains why the spectra of transition probabilities in~\Cref{fig:transampfourier} contain little contribution from higher (unperturbed) energies.
    This can be contrasted with the spectra in~\Cref{fig:transampfouriereq} shown for the case when~$\bket{\Psi_{\rm eq}}$ is chosen as both the initial and final state.
    }
    \label{tab:states}
\end{table*}
\begin{figure*}
    \centering
    \includegraphics[width=0.67\textwidth]{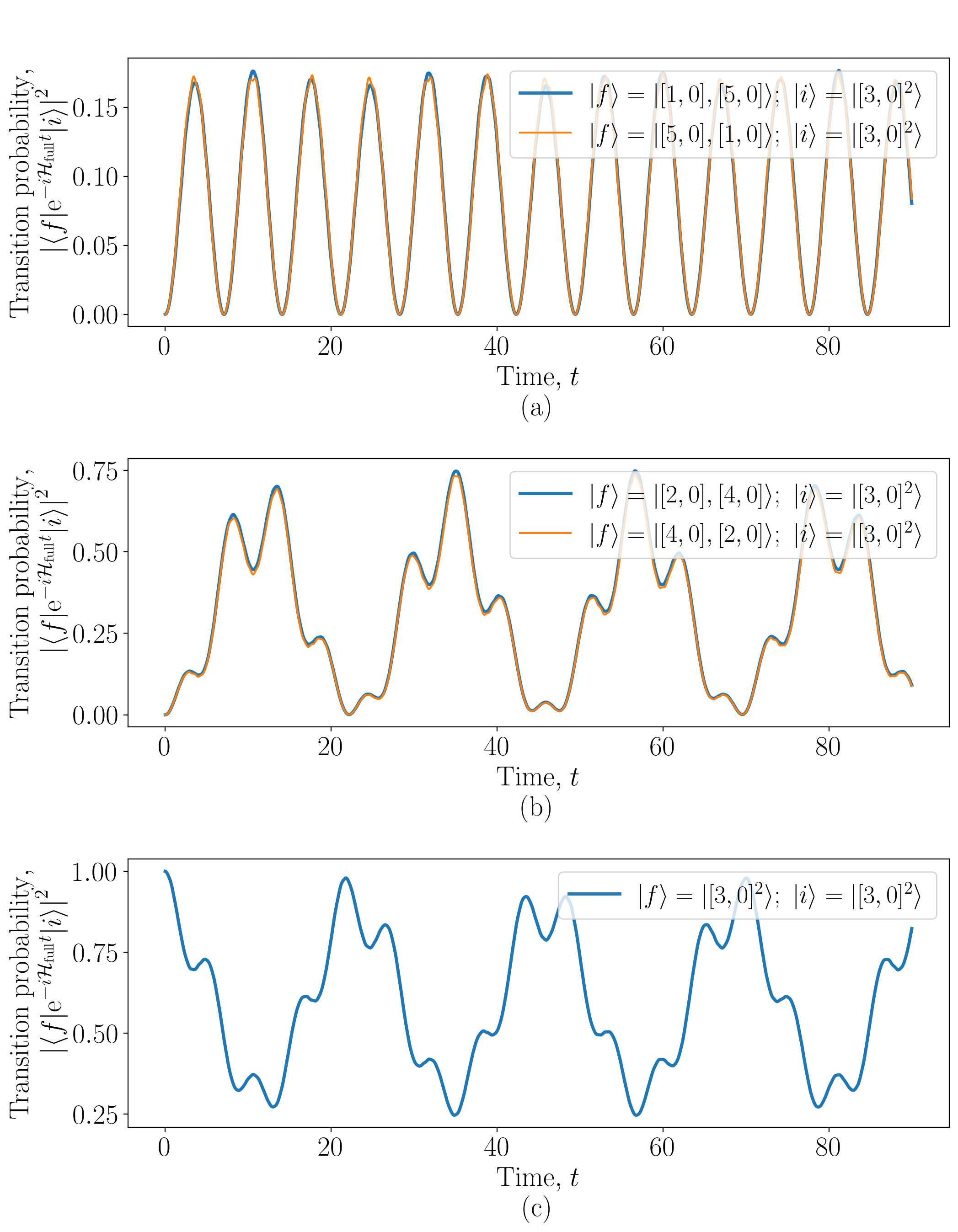}
    \caption{Transition probabilities $\bigl|\sand{f}{\me^{-\iu \Hopfull t}}{i}\bigr|^2$ in the $\phi^4_{1+1}$ theory with
    $\Hopfull$ defined by~\cref{eq:phi4ham}, for the initial state ${\bket{i} = \bket{[3,0]^2}}$ and various final states of the form ${\OPtAd[n,0]\OPtAd[K-n,0]\vac}$.
    As can be seen from their Fourier spectrum (shown is~\Cref{fig:transampfourier}), these probabilities transition probabilities are barely affected by high-energy states of the system.
    }
    \label{fig:transamp}
\end{figure*}

\begin{figure*}[!htb]
    \centering
    \captionsetup{width=.1\linewidth}
    \includegraphics[width=0.67\textwidth]{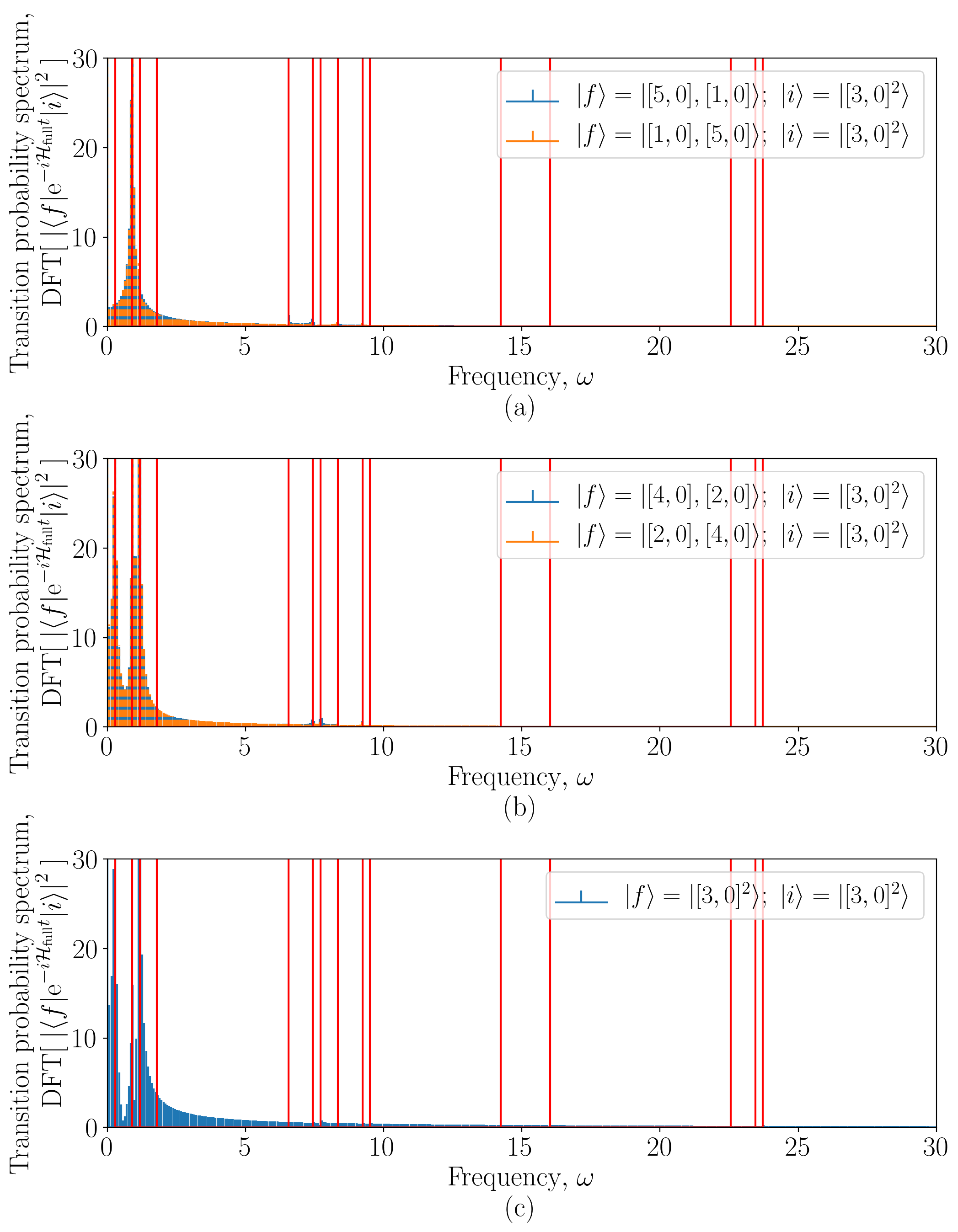}
    \caption{Fourier spectrum of transition probabilities $\bigl|\sand{f}{\me^{-\iu \Hopfull t}}{i}\bigr|^2$ in the $\phi^4_{1+1}$ theory with
    $\Hopfull$ defined by~\cref{eq:phi4ham}, for the initial state ${\bket{i} = \bket{[3,0]^2}}$ and final states $\sket{f}={\OPtAd[6,6-n]\OPtAd[6,n]\vac}$, see~\cref{eq:2K}.
    Red lines indicate the differences ${(E_{\text{full},\,K=6,\,j}-E_{\text{full},\,K=6,\,k})}$ between the energies in the exact spectrum~\eqref{eq:exactspec}.
    Whether certain bands are present or absent in the spectrum, is defined by the form of the corresponding states shown in~\Cref{tab:states}.
    As follows from the figure, the initial and final scattering states have little overlap with eigenstates of $\Hopfull[K=6]$ of high LF energy (see also~\cref{footnote:energydiff}).
    For comparison, see the spectrum of transition probabilities for $\sket{i}=\sket{f}=\bket{\Psi_{\rm eq}}$, with $\bket{\Psi_{\rm eq}}$ being the equal weight superposition of $\Hopfull[K=6]$ eigenstates, which is shown in~\Cref{fig:transampfouriereq}.
    }
    \label{fig:transampfourier}
\end{figure*}

\begin{figure*}[!htb]
    \centering
    \includegraphics[width=0.67\textwidth]{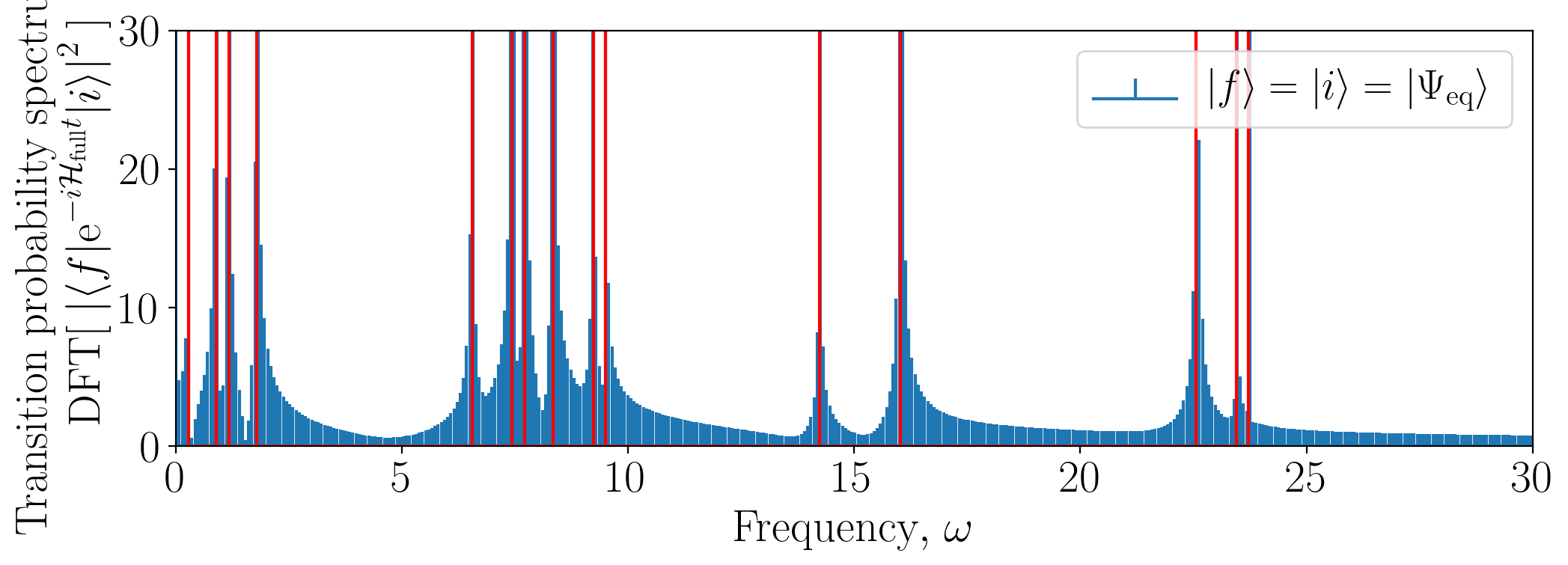}
    \caption{Fourier spectrum of the transition probability $\bigl|\sand{\Psi_{\text{eq}}}{\me^{-\iu \Hopfull t}}{\Psi_{\text{eq}}}\bigr|^2$ for an equal weight superposition of $\Hopfull[K=6]$ even sector eigenstates. The state $\bket{\Psi_{\text{eq}}}$ does not carry physical meaning, the plot is shown to be contrasted with~\Cref{fig:transampfourier}, where the frequencies are shifted toward the left of the spectrum.
    }
    \label{fig:transampfouriereq}
\end{figure*}

Once the initial and final states are determined, the unitary time evolution is simulated using the full Hamiltonian operator acting on momentum modes up to~${K=6}$.
The transition probabilities for the chosen initial and final states are shown in~\Cref{fig:transamp}.
In order to make sense of these graphs, we calculate their Fourier transform (shown in~\Cref{fig:transampfourier}) and confirm that the plots in the frequency domain have their peaks in the points corresponding to the differences ${(E_{\text{full},\,K=6,\,j}-E_{\text{full},\,K=6,\,k})}$  between the eigenvalues of the exact spectrum:\footnote{\label{footnote:energydiff}If
${\me^{-\iu H t}\sket{i} =\sum_n \hspace{-0.5pt} c_n\me^{-\iu\omega_n t}\sket{n}}$ and ${\sket{f} = \sum_m
\hspace{-0.5pt} d_m\sket{m}}$, then ${\abs{\sand{f}{\exp(-\iu H t)}{i}}^2 =  \sum_{mn}d_n^*c_nd_mc_m^*\me^{-\iu(\omega_n-\omega_m) t} }$, where $\omega_n$ and $\sket{n}$ are the eigenvalues and eigenvectors~of~$H$.
}
\begin{equation}
\label{eq:exactspec}
\begin{alignedat}{9}
    &\operatorname{spec} H_{K=6,\,\text{even}} \\
    &\hspace{0.903mm}=
    \{
    0.605,
    0.879,
    1.769,
    8.313,
    10.104,
    24.329
    \} \, .
\end{alignedat}
\end{equation}
To further interpret plots in~\Cref{fig:transampfourier}, it is useful to compare the initial and final scattering states with the eigenvectors of the ${K=6}$ Hamiltonian corresponding to eigenvalues in~\eqref{eq:exactspec}, see~\Cref{tab:states}.
While the Hamiltonian operator used for generating time evolution in~\Cref{fig:transamp} encodes the entire spectrum of the system, the similarity of initial and final states with the lowest states in the ${K=6}$ sector explains why the effects of higher frequencies are barely noticeable on frequency plots in~\Cref{fig:transampfourier}.
These plots can also be contrasted with the transition probability $\bigl|\sand{\Psi_{\text{eq}}}{\me^{-\iu \Hopfull t }}{\Psi_{\text{eq}}}\bigr|^2$ of a trial state $\sket{\Psi_{\text{eq}}}$, an equal superposition of all the six eigenstates from the $K=6$ even sector, see~\Cref{fig:transampfouriereq}.

Let us further investigate the effect of higher momentum modes on the evolution of low momentum states.
To do so, we consider the time-dependent parton distribution function of the initial state, defined as:
\begin{equation}
\label{eq:pdfphieq}
\begin{alignedat}{8}
    \text{PDF}(x=\Moden/K, t) &=
    \sand{i(t)}{\OPad_\Moden\OPa_\Moden}{i(t)}
    \\&=
    \bigl\langle
    i
    \bigl|\bigr.
    \me^{i \Hopfull t}
    \OPad_\Moden\OPa_\Moden
    \me^{-i \Hopfull t}
    \bigl|\bigr.
    i
    \bigr\rangle
    \ .
\end{alignedat}
\end{equation}

\begin{figure}[ht]
    \centering
    \includegraphics[width=0.45\textwidth]{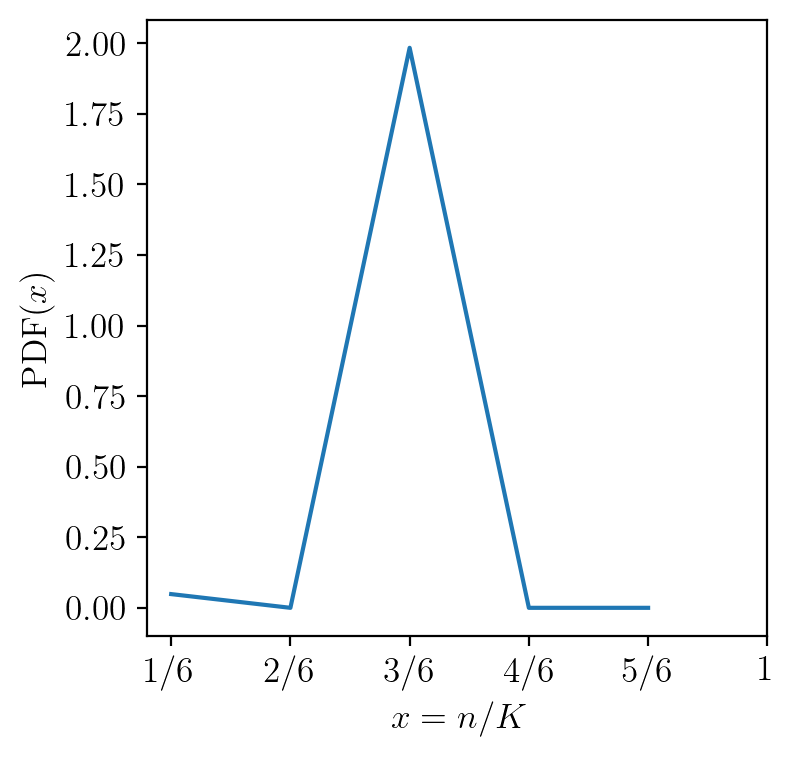}
    \caption{${\text{PDF}(x=n/K, t=0)}$, the parton distribution function, as defined in~\eqref{eq:pdfphieq}, for the initial state ${\sket{i}=\bket{[3,0]^2}}$, as defined in~\cref{eq:2K} and shown in~\Cref{tab:states}.
    }
    \label{fig:pdf_init}
\end{figure}

\begin{figure}
    \centering
    \includegraphics[width=0.45\textwidth]{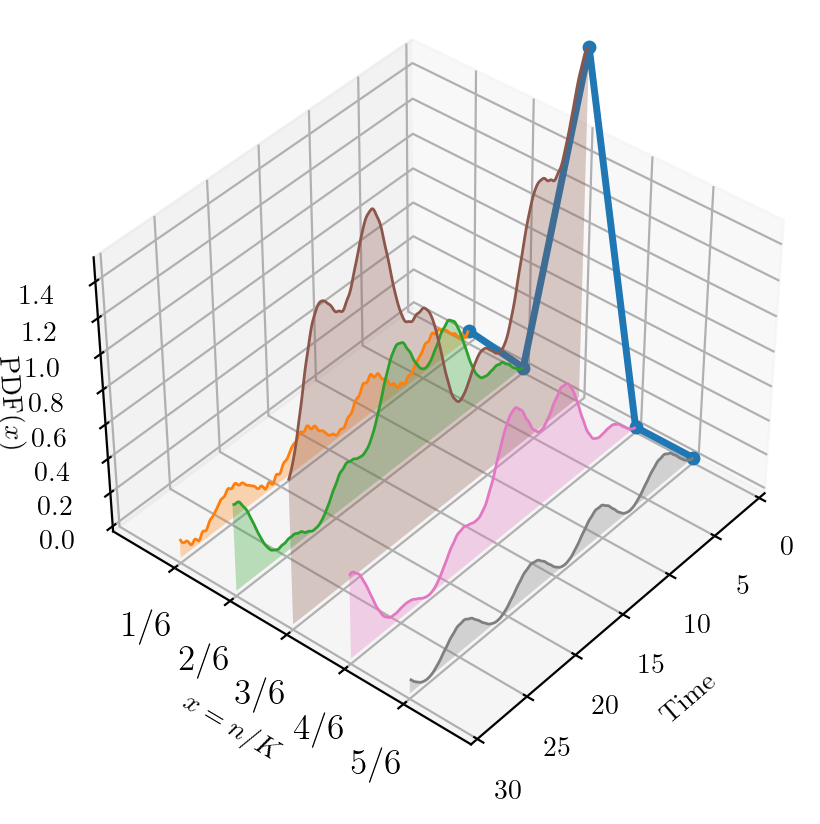}
    \caption{Time-dependent parton distribution function from~\cref{eq:pdfphieq}
    for the state ${\sket{i}=\bket{[3,0]^2}}$.
    The plot illustrates how the the modes of higher LF momentum (${\Moden=4,5}$) participate in the time evolution of the state ${\sket{i}=\bket{[3,0]^2}}$, in which only the modes ${\Moden=1,3}$ were initially occupied.
    }
    \label{fig:pdf_time}
\end{figure}

The initial PDF of the state $\sket{i}=\bket{[3,0]^2}$ is shown in~\Cref{fig:pdf_init}.
As follows from the form of this state (see~\Cref{tab:states}), the PDF is dominated by modes of momentum $1$ and $3$.
The time evolution of this initial PDF is shown in~\Cref{fig:pdf_time}. The plot illustrates the effect of higher-momentum modes $4$ and $5$ on the evolution of the initial state.\footnote{The mode of momentum $6$ is absent in~\Cref{fig:pdf_time} because at harmonic resolution $K=6$ the only state with this mode, $\sket{6}$, contains a single particle and thus belongs to the odd sector. }

In order to calculate the transition probability $\bigl|\sand{f}{\me^{-\iu \Hopfull t}}{i}\bigr|^2$  between the two-particle states in the interacting theory, we defined those using operators ${\OPtAd[n][0]}$ creating the eigenstates of the interacting theory from vacuum.
We constructed such operators with the aid of the wave operators $\Wop[K]$ which relate the eigenbases of the free and interacting theories at particular values of harmonic resolution $K$, see~\cref{eq:OPtad,eq:OPtadvac}.
To find the wave operator $\Wop[K]$, we diagonalized the interacting Hamiltonian matrix in the basis of Fock states, calculated the logarithm of the corresponding modal matrix, and found the simplest cluster operator $\Vop[K]$ of such a form that the unitary transformation ${\Wop[K] = \me^{-\iu \Vop[K]}}$ obeys the desired properties, see~\cref{eq:unit,eq:act,eq:vset}.
We also contrasted the so-defined operators with those obtained via the definition~\cref{eq:defP}, which only involves polynomials in creation operators.
The constructed multi-particle states predominantly belong to the low LF energy subspace of the theory (in the sense of both free and interacting Hamiltonians), yet their time evolution is affected by states of higher LF energies.
While the considered model was simple enough for the calculations to be performed exactly, the practical usage of the method would rely on approximate techniques and/or quantum computing.
In~\Cref{sec:scat_quant}, the latter path is outlined.

\makeatletter\onecolumngrid@push\makeatother
\newpage
\newgeometry{left=2cm,right=2cm,top=1cm,bottom=1cm,headsep=-2cm,bindingoffset=0cm}
\clearpage
\newgeometry{left=2.5cm,right=2.5cm,top=3cm,bottom=3cm,bindingoffset=0cm}
\makeatletter\onecolumngrid@pop\makeatother

\section{Quantum Simulation of Scattering\label{sec:scat_quant}}

The approach to scattering given in~\Cref{sec:dlcq,sec:scatter_prob} is well suited to serve as a starting point for designing quantum simulation algorithms.
Implementing the action of operators, restricted to a certain subset of modes, is achieved via manipulating the corresponding qubit registers on the quantum computer.
If the initial wavefunctions of incoming particles do not overlap, separate state preparation procedures can be used for the disjoint parts of the system.
The corresponding quantum circuits will then act on disjoint subsets of qubits, and the usage of variational and projection-based techniques is possible.
In a more general scenario, when preparation of the composite particles involves overlapping sets of modes, it is most natural to prepare the initial states with the aid of the wave operator.
The action of the latter can be then implemented by means of such techniques as adiabatic state preparation or GWW flow.

\begin{figure*}

\begin{tikzpicture}[baseline={([yshift=-.5ex]current bounding box.center)}]
\centering
\node[scale=1]{
\begin{quantikz}[slice label style ={inner sep=10pt,anchor=south west,rotate=40}, column sep = .2cm, row sep = .4cm]
\lstick{$\sket{n=1}$}
& \gate[5]{U_{[K_1,0],[K_2,0]}}
&\qw
& \midstick[5,brackets=none]{=}
& \gate[5]{\Wopd[K_2]}
& \qw
& \gate[5]{\Wop[K_2]} \slice{}
& \gate[3]{\Wopd[K_1]}
& \qw
& \gate[3]{\Wop[K_1]}
& \qw
\\
\lstick{\ldots} &
&\ghost{.}\qw
&&\ldots&\ldots&\ldots&\ldots&\ldots&\ldots&
\\
\lstick{$\sket{n=K_1}$}
&\qw
& \ghost{.}\qw
&& \qw
& \qw
& \qw
& \qw
& \gate{\OPad_{K_1}}
& \qw
& \qw
\\
\lstick{\ldots}
&\qw&\ghost{.} \qw
&&\ldots&\ldots&\ldots&\ldots&\ldots&\ldots&
\\
\lstick{$\sket{n=K_2}$}
& \qw &\qw
&&\qw
& \gate[1]{\OPad_{K_2}}
& \qw
& \qw
& \qw
& \qw
& \qw
\end{quantikz}
};
\end{tikzpicture}
\caption{Quantum circuit for preparing a two-particle state~\eqref{eq:K1K2}. Single-mode creation operators $\OPad_{K_{1,2}}$ can be implemented using ancillary qubits (see~\Cref{app:creat}). The first $\Wopd[K_2]$ circuit can be omitted if one starts from the vacuum state.}
\label{fig:K1K2}
\end{figure*}
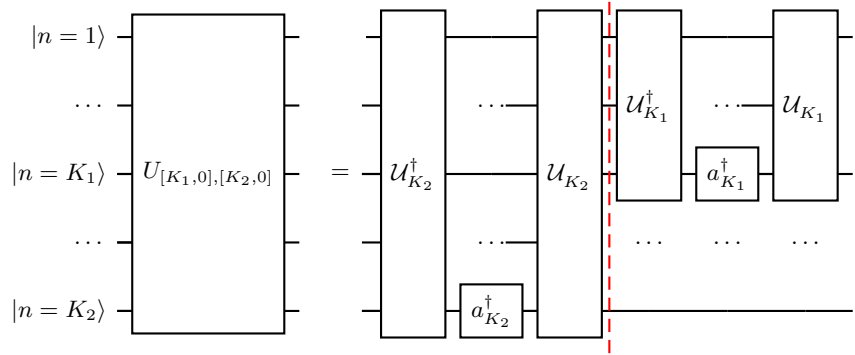

Consider a circuit for preparing the state $\bket{[K_1,0],[K_2,0]}$ defined in~\eqref{eq:K1K2}.
As shown in~\Cref{fig:K1K2}, adding to the state a particle of momentum $K_1$ ($K_2$) only involves acting on modes of momentum up to $K_1$ ($K_2$).
The single-mode creation operators $\OPad_{K_{1,2}}$ can be implemented using ancillary qubits (see e.g. step~2 in Sec.~3.2 of~\cite{2011arXiv1112.4833J} or Section~V. in~\cite{kirby2021quantum}).
The first $\Wopd[K_2]$ gate can be omitted if one starts from a vacuum state.
Thus, the cost of the state preparation circuit comes from implementing wave operator (twice for each particle), as well as from implementing the action of the free creation operator.
In a similar manner one can design circuits for efficient preparation of any states of the form~\eqref{eq:consec2}.

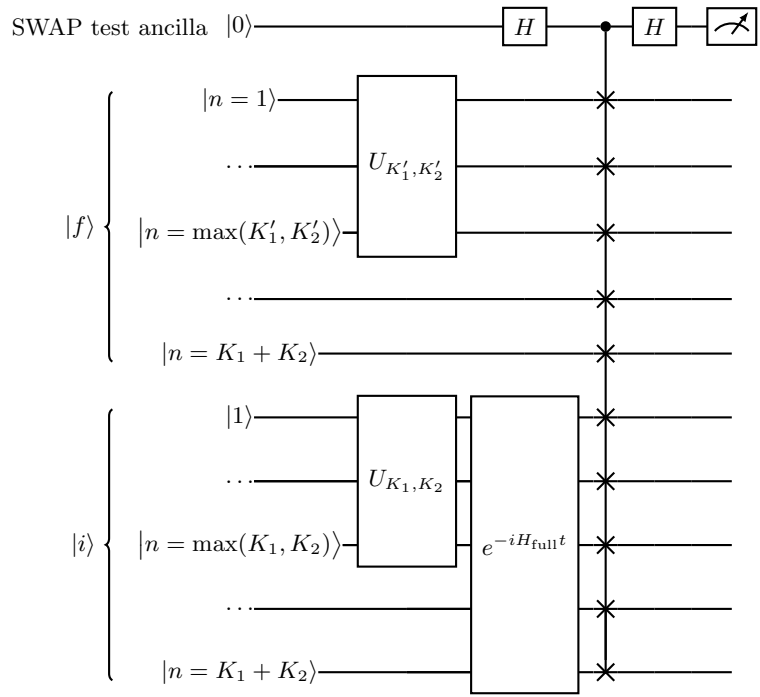
\begin{figure*}
\centering
\begin{tikzpicture}[baseline={([yshift=-.5ex]current bounding box.center)}]
\node[scale=1]{
\begin{quantikz}[slice label style ={inner sep=10pt,anchor=south west,rotate=40}, column sep = .2cm, row sep = .4cm]
\lstick{SWAP te\rlap{st ancilla}}
& \ket{0}                             & \qw & \gate{H} & \ctrl{10} & \gate{H} &  \qw & \meter{} \\
\lstick[wires = 5]{$\ket{f}$}
& \ket{n=1}                           & \gate[3]{U_{K_1^\prime,K_2^\prime}} & \qw  & \targX{} & \qw  & \qw & \qw \\
& \ldots                              & \qw & \qw  & \targX{} & \qw  & \qw & \qw\\
& \ket{n=\max(K_1^\prime,K_2^\prime)} & \qw & \qw & \targX{} & \qw & \qw & \qw\\
& \ldots                              & \qw & \qw  & \targX{} & \qw  & \qw & \qw\\
& \ket{n=K_1+K_2}                     & \qw & \qw  & \targX{} & \qw & \qw & \qw\\
\lstick[wires = 5]{$\ket{i}$}
& \ket{1}                             & \gate[3]{U_{K_1,K_2}} & \gate[5]{e^{-i H_{\text{full}} t}} & \targX{} & \qw & \qw & \qw \\
& \ldots                              & \qw & \qw  & \targX{} & \qw & \qw & \qw\\
& \ket{n=\max(K_1,K_2)}               & \qw & \qw  & \targX{} & \qw & \qw & \qw\\
& \ldots                              & \qw & \qw  & \targX{} & \qw  & \qw & \qw\\
& \ket{n=K_1+K_2}                     & \qw & \qw  & \swap{-1} & \qw & \qw & \qw
\end{quantikz}
};
\end{tikzpicture}
\caption{Quantum circuit for measuring the transition probability~$\bigl|\sand{f}{\me^{-\iu H_{\text{full}}t}}{i}\bigr|^2$ using the SWAP-test.}
\label{fig:swap}
\end{figure*}

Following the state preparation step depicted in~\Cref{fig:K1K2}, one evolves the system in time and measures the observables of interest in the final state of the system.
The simplest (yet likely not the most efficient) way of calculating the transition probability~$\bigl|\sand{f}{\me^{-\iu \Hopfull t}}{i}\bigr|^2$ amounts to doubling the number of qubits and using the SWAP-test~\cite{lin2022lecture} circuit shown in~\Cref{fig:swap}.

While in~\Cref{fig:K1K2,fig:swap} we implicitly assumed the usage of \emph{direct} mapping~\cite{Kreshchuk:2020dla}, in which a particular qubit register is assigned to each momentum mode, the described approach can be implemented using any available encoding and state preparation algorithm, such as that discussed in~\cite{kirby2021quantum}.

Efficient realization of~\cref{eq:defP} containing an exponential number of terms on a quantum computer would pose a challenging task.
Given a state ${\sum_{\FF} c^{\hphantom{\dagger}}_{\FF}\Fock = \sum_{\FF} c^{\hphantom{\dagger}}_{\FF} \mathsf{a}^{\dagger}_{\FF} \vac}$, implementing the action of the operator ${ \sum_{\FF} c^{\hphantom{\dagger}}_{\FF} \mathsf{a}^{\dagger}_{\FF}}$
would likely require access to a subroutine producing the action of $\mathsf{a}^{\dagger}_{\FF}$ controlled on the state $\Fock$.
This would necessitate the usage of fault-tolerant hardware, in which case using the wave operator approach seems to be a more plausible option.
A possible near-term strategy could amount to representing the ground state as a superposition of a polynomial number of basis states using subspace-based methods~\cite{parrish2019quantum,stair2020multireference,yoshioka2022generalized,klymko2022real,seki2021quantum,cortes2022quantum,kirby2023exact}.

\section{Summary and Outlook}

In this work, we proposed a framework for simulating scattering processes, which generalizes existing techniques from non-relativistic many-body quantum mechanics~\cite{szabo1996modern,doi:10.1063/1.477094,C9CP06376E} and relativistic quantum field theory~\cite{varybasis,PhysRevD.101.076016,preskill1,atas20212,Surace_2021} in several  directions.
Our approach is well-suited for the studies of strongly interacting systems, in which the initial and final states are comprised of multiple composite particles with overlapping wavefunctions.
In analogy with the free theory, these states are constructed with the aid of operators creating eigenstates of the interacting  Hamiltonian from vacuum.
Such operators are defined in terms of creation operators of the free theory and wave operators~\cite{kato1966wave,nguyen1985review,evangelista2014driven} relating the eigenbases of the free and interacting theory at particular values of cutoffs, see~\cref{eq:creatop}.
This definition is related to the notion of effective particles in LF QFT~\cite{glazek1993renormalization,glazek1994perturbative,glazek2012perturbative,glazek2012renormalization,glazek2013fermion,Serafin:2019vuk,glazek2021elementary,glazek2017renormalized,glazek2017effective,gomez2017asymptotic} and allows for several ways of implementation.
Note, however, that operators defined in~\cref{eq:creatop} are not the \emph{effective particle operators}~\cite{glazek1993renormalization,glazek1994perturbative,glazek2012perturbative,glazek2012renormalization,glazek2013fermion,Serafin:2019vuk,glazek2021elementary,glazek2017renormalized,glazek2017effective,gomez2017asymptotic} of the combined system in the sense of RGPEP (at infinite flow parameter), as they are defined in theories with smaller cutoffs.
This explains why their successive application does not produce eigenstates of the combined system, and leads to their non-commutativity which has to be further addressed in a future work.

We illustrated the construction of multi-particle states using, as an example, the $\phi^4$ theory in ${1+1\mathrm{D}}$ formulated  in the language of the Discretized Light-Cone Quantization framework~\cite{phi4in2d}.
To construct the wave operator exactly, we used the unitary coupled cluster procedure~\cite{shen2017quantum,Romero_2018,filip2020stochastic}.
We also argued that our approach provides a formulation suitable for simulating scattering of composite particles on quantum computers.
The most natural implementation of the wave operators is via adiabatic state preparation (ASP).
This means that the efficiency of the quantum algorithms is subject to the usual conditions for the efficiency of ASP~\cite{preskill1,aspuru2005simulated,veis2014adiabatic,barends2016digitized,wan2020fast,sugisaki2022adiabatic,ciavarella2023state}.
Quantum algorithms based on the similarity renormalization group are emerging~\cite{gluza2022double}.

While preparation of the wave operator, generally, requires the knowledge of the entire spectrum of the system, of great interest is the possibility of its approximations limited to the low-energy subspace of the Hamiltonian, which can be achieved on near-term quantum hardware using subspace-based methods~\cite{parrish2019quantum,stair2020multireference,yoshioka2022generalized,klymko2022real,seki2021quantum,cortes2022quantum,kirby2023exact}.
We discussed the construction of circuits for initial and final state preparation, as well as measurement procedure, and left for further investigation the construction of circuits implementing  Galilean boosts and LF momentum transformations.

In the present work, we ignored the issue of renormalization.
The LF formulation of QFT provides various ways to perform non-perturbative renormalization of QFT.
An important question is, therefore, which of these methods is most suitable for our approach.
Of particular interest is the possibility of combining the usage of the double commutator flow equation for both renormalizing the canonical QFT and solving it numerically~\cite{evangelista2014driven,li2015multireference,li2016towards,hannon2016integral,li2017driven,li2017low,li2018driven,li2019multireference,wang2019analytic,zhang2019improving,gluza2022double}.

Our work further motivates the development of discretized LF QFT in the position representation.
When choosing a discretization scheme for performing calculations in LF QFT on classical computers, one typically prioritizes the optimal basis choice~\cite{varybasis,li2015ab,quarkonium} allowing for better convergence and/or earlier truncation~--- similar to the studies of localized bound states in quantum chemistry~\cite{szabo1996modern,helgaker2013molecular} and low-energy nuclear physics~\cite{barrett2013ab,binder2016effective}.
Indeed, it is the dimension of Hilbert space and the sparsity of the Hamiltonian matrix that ultimately determine the computational complexity of classical simulation.
On the contrary, the major parameters determining the complexity of quantum simulation are the number of elementary operator terms in the Hamiltonian and their locality.
From this perspective, the optimal choice is given by spatial discretization, which was utilized in the construction of recently developed nearly-optimal algorithms for quantum simulation of quantum chemistry~\cite{babbush2017low,low2018hamiltonian,babbush2019quantum}, and will be investigated in the context of LF QFT.
In addition to these arguments, spatial discretization may provide a way to construct non-overlapping wavefunctions of LF bound states, leading to the factorized form of the initial state, which would enable the usage of variational and projection-based state preparation techniques.

\section{Acknowledgements}

This material is based upon work supported by the U.S. Department of Energy, Office of Science, National Quantum Information Science Research Centers, Quantum Systems Accelerator.
MK acknowledges additional support from the DOE grant PH-HEP24-QuantISED, B\&R KA2401032/34, and is grateful to Stanislaw D. G\l{}azek for fruitful discussions that greatly improved the manuscript. JPV acknowledges support from US Department of Energy grant DE-SC0023692. JPV and PJL acknowledge support from US Department of Energy grant DE-SC0023707.

\newpage

\bibliography{main}
\bibliographystyle{unsrt}

\begin{appendices}
\crefalias{section}{appendix}

\section{The general case\label{sec:definitions}}

In this section, we extend the discussion of states with multiple composite particles from the LF theories in ${1+1\mathrm{D}}$ with harmonic resolution being the only quantum number discussed in~\Cref{sec:dlcq} to more general theories in higher dimensions with multiple conserved charges.
We mainly remain focused on applications to relativistic systems in the LF quantization and non-relativistic systems.
We start off by listing a number of properties, which we assume to hold for a particular discretized formulation of QFT, in order for our following derivations to be applicable.
We begin by assuming the existence of the Hamiltonian operator $\Hopfull$ of the full interacting theory, acting on a countable orthonormal set of basis states $\bigl\{\Fock\bigr\}$.
In QFT, these typically arise in the following digitization schemes~\cite{klco2018digitization,liu2021variational}:
\begin{enumerate}
    \item Field basis in position space.
    \item Field basis in momentum space.
    \item Harmonic oscillator basis in the position space.
    \item Harmonic oscillator basis in the momentum space.
    \item Harmonic oscillator basis in the space of solutions of some single-particle wave
    equation~\cite{varybasis}.
\end{enumerate}
Note that 4 is a particular case of 5, with the equation defining the single-particle basis set being the free field equation whose solutions are the plane waves.

In what follows, we use assume case 5, as it is more suitable for describing complex bound states.
Alternative approaches have been considered in Refs.~\cite{preskill1,preskill2,liu2021variational,Surace_2021}.
In terms of quantum simulation, our choice has two major drawbacks: (a) even for local Lagrangians, the Hamiltonian is no longer geometrically local, which increases either the requirements for the device connectivity or simulation costs.
For example, in the momentum basis, any two momentum modes can interact with one another as long as the overall momentum of  particles involved in the interaction is conserved, and
(b) the number of terms in the Hamiltonian has worse scaling with problem size, as compared to the case of the spatial lattice.

In second-quantized formulations, the number of elementary operator terms in discretized Hamiltonians scales as $O(N^{g})$ for local theories in the plane wave basis or as $O(N^{g-1})$ otherwise, where $N$ is the size of the discretization grid and $g$ is highest power of the interaction term in the Lagrangian, such as $4$ in the $\phi^4$ theory.
At the same time, the number of terms in first-quantized discretized Hamiltonians contains $O(N)$ terms for local theories and $O(N^2)$ terms if non-local interactions are considered (such as Coulomb interaction or effective interactions used in nuclear physics).
Those disadvantages may be outweighed by the fact that the second-quantized formalism often allows for a more efficient description of the system via the usage of highly optimized basis sets, as in the cases of the orbital formulation of quantum chemistry or the nuclear shell model.
Alternatively, one may consider developing quantum simulation algorithms based on the transverse lattice formulation of LF physics~\cite{burkardt1999gauge,bardeen1976local,bardeen1980hadron,burkardt1994light,dalley1999transverse,burkardt1999gauge,dalley2002hadrons,burkardt2002relativistic,burkardt2001study,burkardt2002study,seal2002investigating,chakrabarti2003fermions,Chakrabarti:2003wi}
In this scenario, the number of terms in the Hamiltonian would scale as $O(K^{g-1} N_\bot)$, which is an improvement over $O(N^{g-1})= O(K^{g-1} N_\bot^{g-1})$ in DLCQ or $O(K^{g-1} N_\bot^{g})$ in BLFQ.
Of great interest are other continuous-space basis sets, such as
wavelets~\cite{altaisky2010quantum,altaisky2013continuous,bulut2013wavelets,altaisky2020wavelet,bagherimehrab2023efficient}, --  and so is the possibility of efficient switching over different bases throughout the quantum computation, in order to take advantage of both the
Hamiltonian locality and a convenient basis choice~\cite{barata2021single}.

We assume that the second-quantized Hamiltonian $\Hopfull$ can be written in terms of creation and annihilation operators acting on \emph{Fock states}~$\bigl\{\Fock\bigr\}$, the eigenstates of another second-quantized Hamiltonian operator~$\Hopfree$.
Following the usual convention, we call the latter Hamiltonian ``free'', which in the present context is no more than a label for the Hamiltonian (typically exactly-solvable) whose basis is used for solving~$\Hopfull$.
In cases when gauge fields are present in the theory, we further assume that a particular gauge is chosen and fixed~\cite{Brodsky:1997de}.

Next, we make an assumption that the free discretized QFT admits a finite Canonical Set of Commuting Observables (CSCO):
\begin{equation}
\label{eq:CSCOfree}
\begin{alignedat}{99}
\text{Free CSCO}&\!:\quad&&\{\Hopfree,\,
    &&\chargeop[1],\,
    \chargeop[2],\,\ldots\} ,\,\\
&&&
\hspace*{-1.7cm}
\mathrlap{
[\Hopfree,\,\chargeop[k]] =
[\chargeop[j],\,\chargeop[k]] = 0\,.
}
\end{alignedat}
\end{equation}
The elements of CSCO, other than the Hamiltonian, will be termed \emph{charges}.
Those will typically include such operators as momentum, angular momentum, electric charge, and other Noether charges.\footnote{This requirement is not entirely trivial, as in a (continuous) relativistic QFT, the algebra of observables associated with a bounded region of spacetime may be a type-III von Neumann algebra, which cannot have a CSCO in any representation on a separable Hilbert space~\cite{YNGVASON2005135}.
Moreover, pathological counterexamples may be constructed in finite-dimensional spaces as well~\cite{Yngvason:2004uh,beisbart2011probabilities}.
}

In order to explicitly construct $\Hopfree$, one introduces a single-particle Hamiltonian eigenvalue equation, whose solutions are labeled with a multi-index~${\Modecollective=\{\Modecollective[1],\,\Modecollective[2],\ldots\}}$ storing the eigenvalues of charges $\{\chargeop[1],\,\chargeop[2],\,\ldots\}$.
Note that in some cases the charge operator $\chargeop[j]$, corresponding to the quantum number $\Modecollective[j]$, is not known (even though in quantum mechanics it is always possible to \emph{formally} define a symmetry operator for any quantum number, see~\Cref{app:charges}).
This, however, does not pose a problem, as one can consider such quantum numbers as additional labels for states with fixed values of known charges.
The set of all possible assignments of quantum numbers for the single-particle Hamiltonian equation is denoted by~${\Modeset\ni\Modecollective}$.
It can be made finite by imposing a cutoff on allowed eigenvalues of charge operators (typically those related to momentum).

The operator $\Hopfree$ acquires the form
\begin{equation}
    \label{eq:Hopfreeaa}
    \Hopfree = \sum_{\Modecollective\in\Modeset} \Efree_{\Modecollective} \OPad_{\Modecollective} \OPa_{\Modecollective} \, ,
\end{equation}
where the dependence of $\Efree_{\Modecollective}$ on $\Modecollective$ encapsulates the dispersion relation of the free theory, while operators on the RHS of~\cref{eq:Hopfreeaa} are assumed to be either bosonic or fermionic (consideration of exotic particles such as anyons may require special treatment and is left for future work).
Operators~$\OPad_{\Modecollective}$ create excitations  acting upon the unique vacuum of the theory, $\vac$.

The single-particle states of the free theory have the form
    $
    \sket{\collective}={\OPad_{\collective}\vac}
    $,
while a general multi-particle Fock state $\Fock$ can be written as
\begin{equation}
    \label{eq:fockstate}
    \Fock = \sket{\collective_1^{\occ_1},\,\collective_2^{\occ_2},\ldots,\,\collective_{\FockNModes}^{\occ_\FockNModes}} \, ,
\end{equation}
where the modes $\collective_j\in\Modeset$ correspond to the solutions of the free single-particle equation, $J$ is the total number of occupied modes in a state, $\occ_j$ denotes the (non-zero) occupancy of the mode~$\collective_j$, and unoccupied modes are omitted.\footnote{$\collective$ denotes a particular mode in a Fock state, while $\Modecollective$ is an index running over all the possible modes from $\Modeset$.
}
\Cref{eq:fockstate} is written in the so-called \emph{compact} notation (typically used in QFT), which shows the occupied modes only.
Alternatively, one could use the \emph{direct} notation (typically used in quantum chemistry) and explicitly indicate the occupancies of all the modes from $\Modeset$.
These notations correspond to various ways of encoding Fock states in the quantum computer~\cite{Kreshchuk:2020dla}.

The occupancies $\occ_j$ take their values in the range
\begin{equation}
    \occ_j \in [1,\,\occ_j^{\text{max}}] \,.
\end{equation}
We store the mode occupancy cutoffs~$\occ_j^{\text{max}}$ as a positive-integer-valued function $\occ_j^{\text{max}} = \occmax(\collective_j)$ on the set~$\Modeset$, given by the structure of the theory and/or imposing cutoffs by hand:
\begin{equation}
    \label{eq:WWW}
    \occmax(\Modecollective):
    \,\,
    \Modecollective\!\in\!\Modeset \mapsto
    \!
    \left\{
    \begin{alignedat}{8}
        &\occmax = 1 &&\mathllap{\text{(fermionic
        $\Modecollective$)}},\\
        &\occmax \in 1,2,3,\ldots\,\, \text{(bosonic
        $\Modecollective$)}&&.
    \end{alignedat}
    \right.
\end{equation}
Lastly, we denote the maximum number of occupied modes in a Fock state by~$\maxmodes$:
\begin{equation}
    \label{eq:maxmodes}
    \FockNModes \leq \maxmodes \, .
\end{equation}

The set of all Fock states with modes in~$\Modeset$ and cutoffs $\occmax(\Modeset)$ and $\maxmodes$ is denoted by ${\FockAll(\Modeset, \occmax,\maxmodes)}$ and termed \textit{Fock basis} associated with ${(\Modeset, \occmax,\maxmodes)}$:
\begin{gather}
\label{eq:LFGWI}
\begin{alignedat}{8}
    \FockAll(\Modeset, &\occmax,\maxmodes) &&= \bigl\{
    \sket{\collective_1^{\occ_1},\,\collective_2^{\occ_2},\ldots,\,\collective_{\FockNModes}^{\occ_J}}\\
    &\bigl|\bigr.
    \collective_j\in&&\Modeset,
    \text{ and }
    \occ_j
    = \occmax(\collective_j),
    \text{ and }
    \FockNModes \leq I
    \bigr\}
    \\&
    &&\equiv
    \bigl\{
    \Fock[\Modeset, \occmax,\maxmodes][][]
    \bigr\}
    \, .
\end{alignedat}
\end{gather}

For two Fock bases $\FockAll(\Modeset_1, \occmax_1,\maxmodes_1)$ and $\FockAll(\Modeset_2, \occmax_2,\maxmodes_2)$, the following holds:
\begin{equation}
\label{eq:fockrestr}
    \left.
    \begin{alignedat}{9}
        \Modeset_1\subseteq\Modeset_2
        \\
        \Modecollective\!\in\!\Modeset_1\!\!:\,\,W_1(\Modecollective)\! \leq\! W_2(\Modecollective)
        \\
        I_1\leq I_2
    \end{alignedat}
    \right\}
    \!\Rightarrow
    \begin{alignedat}{8}
    \FockAll&(\Modeset_1, \occmax_1,\maxmodes_1) \\
    &\!\!\!\!\subseteq
    \FockAll(\Modeset_2, \occmax_2,\maxmodes_2) \, ,
    \end{alignedat}
\end{equation}
\Cref{eq:fockrestr} states the conditions under which one Fock basis is a restriction of another.
If this is the case, any theory defined on ${\Span\FockAll(\Modeset_1, \occmax_1,\maxmodes_1)}$ is also naturally defined on ${\Span\FockAll(\Modeset_2, \occmax_2,\maxmodes_2)}$, as the former Hilbert space is a subspace of the latter.
The sufficient conditions for the number of Fock states to be finite are that $\Modeset$, $\occmax(\Modecollective)$, and $I$ are all finite.

We use~${\statecharge=\{\statecharge[1],\,\statecharge[2],\ldots\}}$ to denote a particular assignment of eigenvalues of charge operators from
~$\{\chargeop[1],\,\chargeop[2],\,\ldots\}$,
acting on a multi-particle Fock state.
The set of all possible charge assignments, which a Fock state from $\FockAll(\Modeset, \occmax,\maxmodes)$ may have, is denoted by~${\Chargeset(\Modeset,\occmax,\maxmodes)}$ and, generally, ${\Modeset\subseteq\Chargeset}$.
The subset of~${
\FockAll(\Modeset, \occmax,\maxmodes)}$ with charge eigenvalues~$\statecharge$ is denoted by~${
\FockAll[\statecharge](\Modeset, \occmax,\maxmodes)}$:
\begin{equation}
\label{eq:informal}
\begin{alignedat}{9}
    \FockAll[\statecharge]&(\Modeset, \occmax,\maxmodes)
    \\=&
    \bigl\{
    \Fock\in \FockAll(\Modeset, \occmax,\maxmodes)
    \,
    \bigl|\bigr.\textstyle
    \sum_j\occ_j\collective_j = \statecharge
    \bigr\}
    \\\equiv&
    \bigl\{
    \Fock[\Modeset, \occmax,\maxmodes][\statecharge][]
    \bigr\}
    \,.
\end{alignedat}
\end{equation}
Such subsets form bases for the blocks of both free and full Hamiltonians, corresponding to quantum numbers~$\statecharge$.

In certain cases, we shall explicitly specify with the subscript $\Modeset$ the set of modes, upon which a particular second-quantized operator $\mathcal{O}$ acts:
\begin{equation}
    \mathcal{O}_\Modeset=\poly(\OPa_{\Modecollective},\OPad_{\Modecollective})
    \,,\quad
    \Modecollective\in\Modeset \, .
\end{equation}
For an operator $\mathcal{O}$, whose action is originally defined on a set of modes different from~$\Modeset$, its restricted version~$\mathcal{O}_\Modeset$ is obtained by removing all the terms acting on modes not belonging to~$\Modeset$. Note that this is not equivalent to a formal requirement that the action of $\mathcal{O}$ and $\mathcal{O}_\Modeset$ coincide on $\FockAll(\Modeset, \occmax,\maxmodes)$, as one can construct infinitely many second-quantized operators acting identically on $\FockAll(\Modeset, \occmax,\maxmodes)$, by using polynomials of increasing degrees.
We use this idea in~\Cref{sec:dlcq} in the construction of the coupled cluster operator.

At this point, we introduce a slightly unusual notation~-- by~$\OOPad_{\FF}$ we denote the operator creating Fock state~$\Fock$ from vacuum~$\vac$:
\begin{equation}
    \label{eq:fromfreevac}
    \Fock = \OOPad_{\FF} \vac \, .
\end{equation}
Note that~\cref{eq:fromfreevac} generalizes the usual definition of single-particle creation operators, as $\OOPad_{\FF}$ may also create multi-particle states.
In the latter case, $\OOPad_{\FF}$ is a product of single-particle creation operators, with a numerical prefactor ensuring the appropriate normalization of~$\Fock$, see, e.g., operators creating the ground states of $\phi^4$ theory in~\Cref{tab:GSs}.

For simplicity, in the remaining part of the paper we shall assume that the CSCOs of the free and interacting theories are both known and differ only by their Hamiltonian operators, while a more general scenario
is discussed in~\Cref{app:charges}:
\begin{equation}
\label{eq:CSCOfull}
\begin{alignedat}{99}
\text{Full CSCO}&\!:\quad&&\{\Hopfull,\,
    &&\chargeop[1],\,
    \chargeop[2],\,\ldots\} ,\,\\
&&&
\hspace*{-1.7cm}
\mathrlap{
[\Hopfull,\,\chargeop[k]] =
[\chargeop[j],\,\chargeop[k]] = 0\,.
}
\end{alignedat}
\end{equation}

By~$\FFockAll(\Modeset, \occmax,\maxmodes)$ we denote the set of orthonormal simultaneous eigenstates of the full CSCO~\eqref{eq:CSCOfull}.
Furthermore, we assume that
\begin{equation}
    \label{eq:spanspan}
    \Span \FFockAll(\Modeset, \occmax,\maxmodes)
    = \Span \FockAll(\Modeset, \occmax,\maxmodes) \,,
\end{equation}
i.e. that the solutions of the free and interacting theories belong to the same Hilbert space.
Moreover, we assume that each vector~$\FFock\in\FFockAll(\Modeset, \occmax,\maxmodes)$ can be obtained from some element of the Fock basis~$\Fock\in\FockAll(\Modeset, \occmax,\maxmodes)$ via the application of a unitary operator~$\Wop(\Modeset, \occmax,\maxmodes)$:
\begin{equation}
\label{eq:fff}
\begin{gathered}
    \forall\FFock\in\FFockAll(\Modeset, \occmax,\maxmodes)
    \
    \exists
    \Fock\in\FockAll(\Modeset, \occmax,\maxmodes)
    :\\
    \Wop(\Modeset, \occmax,\maxmodes)
    \Fock = \FFock \, ,
\end{gathered}
\end{equation}
which we also informally (element-wise) write as
\begin{equation}
    \label{eq:adiabset}
    \Wop(\Modeset, \occmax,\maxmodes)
    \FockAll(\Modeset, \occmax,\maxmodes) = \FFockAll(\Modeset, \occmax,\maxmodes)\,.
\end{equation}

\Cref{eq:adiabset} is more than a mere observation that any two orthonormal bases in the Hilbert space are related by a unitary matrix.
It states that such a transformation can be given by an operator $\Wop(\Modeset, \occmax,\maxmodes)$ defined in terms of its action on modes from~$\Modeset$.
Within the unitary coupled cluster approach, $\Wop(\Modeset, \occmax,\maxmodes)$ is sought as an exponential of a polynomial in creation and annihilation operators.
Potentially more efficient ways of constructing $\Wop(\Modeset, \occmax,\maxmodes)$ include adiabatic state preparation (which only works in the absence of phase transitions) or the usage of G\l{}azek-Wilson-Wegner double commutator flow equation~\cite{glazek1993renormalization,glazek1994perturbative,wegner1994flow}.
If $\Wop(\Modeset, \occmax,\maxmodes)$ is implemented as an adiabatic interaction turn-on, the states~$\FFockAll(\Modeset, \occmax,\maxmodes)$ are the eigenvectors of~$\Hopfull$ with eigenvalues from the spectrum of~$\Hopfree$.
We also assume the vacuum state, which we assume to be the same for the free and interacting theories (see also~\cref{footnote:vacua}), to be invariant under the action of $\Wop(\Modeset, \occmax,\maxmodes)$:
\begin{equation}
    \label{eq:vacvac}
    \Wop(\Modeset, \occmax,\maxmodes) \vac = \vac \, .
\end{equation}

By~$\FFockAll[\statecharge](\Modeset, \occmax,\maxmodes)$ we denote the subset of $\FFockAll(\Modeset, \occmax,\maxmodes)$ corresponding to a particular choice of~$\statecharge$.
The adiabatic turn-on is expected to commute with the charge operators,
and, therefore, it acts on the Fock space block-wise:
\begin{equation}
    \label{eq:adiab_block_u}
    \Wop(\Modeset, \occmax,\maxmodes)\, \FockAll[\statecharge](\Modeset, \occmax,\maxmodes) =
    \FFockAll[\statecharge](\Modeset, \occmax,\maxmodes)
    \,.
\end{equation}

We are now in position to make the key definition of this section, and introduce an operator $\OOPAd_{\GG}$ creating the eigenstates of the full Hamiltonian operator $\Hopfull$ from the vacuum:
\begin{equation}
    \label{eq:Adef}
    \OOPAd_{\GG} \equiv \Wop(\Modeset, \occmax,\maxmodes) \OOPad_{\FF}\, \Wopd(\Modeset, \occmax,\maxmodes) \, ,
\end{equation}
where the explicit dependence of~$\OOPAd_{\GG}$ on $\Modeset$, $\occmax$, and $\maxmodes$ has been suppressed.

In order to justify the definition~\eqref{eq:Adef}, let us apply $\OOPAd_{\GG}$ to the vacuum:
\begin{equation}
\begin{alignedat}{84}
    \OOPAd_{\GG} \vac
    &=\Wop(\Modeset, \occmax,\maxmodes) \OOPad_{\FF} \, \Wopd(\Modeset, \occmax,\maxmodes)
    \vac
    \\&=
    \Wop(\Modeset, \occmax,\maxmodes) \OOPad_{\FF} \vac
    \\&
    = \Wop(\Modeset, \occmax,\maxmodes) \Fock
    \mathrlap{= \FFock \, .}
\end{alignedat}
\end{equation}

In what follows, we shall often explicitly indicate upon which sets of modes various operators act (e.g., $\Hopfree[\Modeset]$, $\Hopfull[\Modeset]$),
and consider their action on states
of specific charges~$\statecharge$.
Note that the actions of operators $\mathcal{O}_{\Modeset}$ and  $\mathcal{O}_{\Modeset^\prime}$ from theories defined on $\Modeset$ and $\Modeset^\prime\subset\Modeset$ coincide on $\Modeset^\prime$ but generally differ on~$\Modeset$:
\begin{subequations}
\label{eq:adiab_block}
\begin{alignat}{99}
\label{eq:adiab_block_neq}
&
\begin{alignedat}{99}
\mathcal{O}_{{\Modeset}}
    \FockAll[\statecharge](\Modeset^\prime, \occmax,\maxmodes)
    &=
    \mathcal{O}_{{\Modeset^\prime}}
    \FockAll[\statecharge](\Modeset^\prime, \occmax,\maxmodes)
    \\&=
    \FFockAll[\statecharge](\Modeset^{\mathrlap\prime}, \occmax,\maxmodes)
    \,,
\end{alignedat}\\
    &\mathcal{O}_{\Modeset^\prime}
    \FockAll[\statecharge](\Modeset, \occmax,\maxmodes)
    \neq
    \mathcal{O}_{\Modeset}
    \FockAll[\statecharge](\Modeset, \occmax,\maxmodes) \, ,
    \\
&
\hspace{2.72cm}
\mathllap{\Modeset^\prime}\subset \Modeset \, .\nonumber
\end{alignat}
\end{subequations}

The main value of the definition~\eqref{eq:Adef} is in that it allows one to define multi-particle states in the full interacting theory.
Let
\begin{equation}
    \Fock[\Modeset_k, \occmax_k,\maxmodes_k][\statecharge_k][k] \equiv
    \Fock[][][k]\in\FockAll_{\statecharge_k}(\Modeset_k, \occmax_k,\maxmodes_k)
\end{equation}
be some Fock states,  defined on different sets of modes and with different cutoffs.
Let ${\FFock[][][k]=\Wop(\Modeset_k, \occmax_k,\maxmodes_k)\Fock[][][k]}$ be the eigenstates in the in the corresponding interacting theories.
We define a multi-particle state in the interacting theory with $\occint_k$ particles of type ${\FFock[k]_{\statecharge_k}}$ as follows:\footnote{We omit the tildes whenever possible, as the definition~\eqref{eq:mult_int} will only be used for constructing states in the interacting theory.}
\begin{equation}
\begin{alignedat}{8}
    \label{eq:mult_int}
    &\hspace{-1cm}
    \bigl|\GG_1^{\occint_1},\,\GG_2^{\occint_2},\,\ldots\bigr\rangle_{\statecharge}
    \\
    \equiv \ &\Dconst_{\{[\GG_1]^{\occint_1},[\GG_2]^{\occint_2},\ldots\}}
    \\
    &\times \bigl(\OOPAd_{\GG_1}\bigr)^{\occint_1}
    \bigl(\OOPAd_{\GG_2}\bigr)^{\occint_2}\ldots\vac \, ,
\end{alignedat}
\end{equation}
where
\begin{gather}
    \begin{multlined}
    \statecharge
    \equiv
    \{\statecharge[1],\,\statecharge[2],\ldots\}
    \\=
    \sum_k \occint_k\statecharge_k
    \equiv
    \sum_k \occint_k\{\statecharge[1],\,\statecharge[2],\ldots\}_k
    \, ,
\end{multlined}
\\
\OOPAd_{\GG_k} \equiv
\Wop(\Modeset_k, \occmax_k,\maxmodes_k) \OOPad_{\FF_k}\, \Wopd(\Modeset_k, \occmax_k,\maxmodes_k)
\, ,
\end{gather}
and the constant $\Dconst_{\{[\FF_1]^{\occint_1},[\FF_2]^{\occint_2},\ldots\}}$ ensures the appropriate normalization.
The so-defined state
belongs to a theory defined on an enlarged set of modes ${\Modeset=\Modeset_1\cup\Modeset_2\cup\ldots}$ and appropriately extended cutoffs $\occmax(\Modeset)$ and $\maxmodes$.

\begin{figure}
    \centering
    \begin{tikzpicture}
    \filldraw[fill=black, opacity=0.2];

    \draw[line width = 0.3mm] \firstellip node [label={[xshift=0.543cm, yshift=.01cm, rotate=40]$\Span\bigl(\FockAll(\Modeset_2, \occmax_2,\maxmodes_2)\bigr)$}] {};
    \draw[line width = 0.3mm] \secondellip node [label={[xshift=-0.11cm, yshift=-0.17cm, rotate=-40]$\Span\bigl(\FockAll(\Modeset_1, \occmax_1,\maxmodes_1)\bigr)$}] {};
   \draw[line width = 0.3mm] \thirdellip node [label={[xshift=0.0cm, yshift=1.3cm]$\Span\bigl(\FockAll(\Modeset=\Modeset_1\cup\Modeset_2, \occmax,\maxmodes)\bigr)$}] {};

\end{tikzpicture}
    \caption{Hilbert spaces spanning the sets of Fock states ${\FockAll(\Modeset_1, \occmax_1,\maxmodes_1)}$,
    ${\FockAll(\Modeset_2, \occmax_2,\maxmodes_2)}$, ${\FockAll(\Modeset=\Modeset_1\cup\Modeset_2, \occmax,\maxmodes)}$, which are defined on the sets of modes $\Modeset_1$, $\Modeset_2$,
    and $\Modeset=\Modeset_1\cup\Modeset_2$, correspondingly.
    Adding to the system a particle, represented by a state with modes in $\Modeset_k$,  enlarges the number of available Fock states and  expands the Hilbert space of the system.}
    \label{fig:hilbert_spaces}
\end{figure}

Intuitively, the construction in~\cref{eq:mult_int} describes a scenario, in which several composite particles (eigenstates of the corresponding interacting Hamiltonians) are prepared separately and then brought into contact within a combined system, as schematically shown in~\Cref{fig:particle_cartoon}.
The resulting state belongs to the theory with a larger number of modes and/or higher values of mode cutoffs, which can be interpreted as increasing the range of accessible energies, as compared to theories describing individual particles.
While the eigenstates~$\FFock[][][k]$, describing individual particles, belong to the Hilbert spaces~${\Span\FockAll_{\statecharge_k}(\Modeset_k, \occmax_k,\maxmodes_k)}$, the state in~\cref{eq:mult_int} belongs to the tensor product of those:
\begin{equation}
    \Span\FockAll_{ \statecharge}(\Modeset, \occmax,\maxmodes)
    = \bigotimes_k
    \Span\FockAll^{\otimes\occint_k}_{\statecharge_k}(\Modeset_k, \occmax_k,\maxmodes_k) \, .
\end{equation}

The non-trivial outcome of ``bringing the interacting particles together'' is reflected by the fact that states defined in~\eqref{eq:mult_int} are \emph{not} the eigenstates of the interacting theory they belong to.
Indeed, as $\Wop(\Modeset, \occmax,\maxmodes)$ was never used in the definition of the state~\eqref{eq:mult_int},~the latter is not an eigenstate
of the full Hamiltonian~$\Hopfull[\Modeset]$.
Therefore, it undergoes non-trivial time evolution under the action of~$\Hopfull[\Modeset]$.

States of the form~\eqref{eq:mult_int} can be used as initial and final states in scattering processes.
The definition~\eqref{eq:mult_int} is well-suited for quantum simulation, since one can naturally implement adiabatic interaction turn-on restricted to a particular set of modes~\cite{Aharonov:2003:AQS:780542.780546,aspuru2005simulated,du2010nmr,2011arXiv1112.4833J,veis2014adiabatic,barends2016digitized,wan2020fast,sugisaki2022adiabatic,ciavarella2023state}.

\section{Multi-hadron wave functions on the light front\label{sec:lfmanybody}}

Within the LF approach to relativistic dynamics, introduced in~\Cref{sec:dlcq}, the central object of study is the mass-squared operator ${\mathcal{M}^2 = \mathcal{P}^+ \mathcal{P}^- - \bm{\mathcal{P}}^2_\perp}$, where $\mathcal{P}^+$, $\mathcal{P}^-$, and $\bm{\mathcal{P}}^2_\perp$ are the operators of LF longitudinal momentum, LF energy, and transverse momentum, correspondingly.
All of these operators are related to a particular inertial reference frame with coordinates $x^\mu=(x^0, x^1, x^2, x^3)$.
The LF coordinates associated with this reference frame are obtained via a general coordinate transformation
\begin{equation}
\label{eq:etlf}
    \tilde{x}^\mu=(x^+ = x^0 + x^3 , x^- = x^0 - x^3,x^1, x^2 ) \,.
\end{equation}

The wave function of a bound state with momentum ${(P^+, \bm{P}_\perp)}$ (where $P^+$ and $\bm{P}_\perp$ denote the eigenvalues of the $\mathcal{P}^+$ and $\bm{\mathcal{P}}^2_\perp$ operators, correspondingly) can be expressed in terms of single-particle momenta $p^+_i$ and $\bm{p}_{\perp i}$, where
\begin{equation}
    \sum_i p^+_i = P^+ \,,\quad
    \sum_i \bm{p}_{\perp i} = \bm{P}_\perp \,.
\end{equation}
This choice of momenta is tied to particular LF coordinates $\tilde{x}^\mu$ and, consequently, to the coordinates ${x}^\mu$ in the original inertial reference frame.

A great advantage of the LF formalism stems from the ability to use the \emph{intrinsic frame}~\cite{Brodsky:1997de}, in which the wave function of a bound state is expressed in terms of \emph{relative} longitudinal ($x_i$) and transverse ($\bm{k}_{\perp i}$) momentum coordinates defined as
\begin{equation}
    \label{eq:relcoord}
    x_i = {p^+_i}/{P^+} \,, \quad
    \bm{k}_{\perp i} = \bm{p}_{\perp i} - x_i \bm{P}_{\perp} \,,
\end{equation}
so that
\begin{equation}
    \label{eq:relcoordmom}
    \sum_i x_i = 1 \,,\quad
    \sum_i \bm{k}_{\perp i} = 0\,.
\end{equation}
These coordinates provide a boost-invariant description of hadron's internal structure, independent of its center-of-mass motion.\footnote{Note that the factorizability of a relativistic wavefunction into the parts, corresponding to center-of-mass and internal motion, is a unique feature of light-cone quantization, since the notion of center of mass is, generally, not frame-invariant.}

The wavefunction describing a system of multiple hadrons can be prepared in the following way:
\begin{enumerate}
    \item The wavefunction $\widebar{\Psi}^{(l)} (\widebar{x}_i, \widebar{\bm{k}}_{\perp i})$ of $l$th hadron  state is found in its intrinsic frame.
    We shall assume that operators $\widebar{\OOPA}{}^{(l)\dagger}$ such that
    \begin{equation}
        \sket{\widebar{\Psi}^{(l)} (\widebar{x}_i, \widebar{\bm{k}}_{\perp i})} = \widebar{\OOPA}{}^{(l)\dagger} \vac
    \end{equation}
    are known.

    \item Given the LF momenta of hadrons ${(P^{(l)+}, \bm{P}^{(l)}_\perp)}$, specified in a particular reference frame, the boosted wavefunctions $\Phi^{(l)} ({p}^+_i, {\bm{p}}_{\perp i})$ are calculated, where
    \begin{equation}
    \label{eq:boost1}
    \widebar{x}_i = {{p}^+_i}/{P^{(l)+}} \,, \quad
    \widebar{\bm{k}}_{\perp i} = {\bm{p}}_{\perp i} - \widebar{x}_i \bm{P}^{(l)}_{\perp} \,.
    \end{equation}
    These wavefunctions, in turn, are obtained from vacuum as
    \begin{equation}
        \sket{\Phi^{(l)} ({p}^+_i, {\bm{p}}_{\perp i})} = \mathcal{A}^{(l)\dagger} \vac \,,
    \end{equation}
    where
    \begin{equation}
    {\mathcal{A}^{(l)\dagger} = \mathcal{B}^{(l)}\widebar{\OOPA}{}^{(l)\dagger}\mathcal{B}^{(l)\dagger}}
    \end{equation}
    with $\mathcal{B}^{(l)}$ implementing the transformation~\eqref{eq:boost1}.

    \item The wavefunctions of hadrons $\Psi^{(l)}(x_i,\bm{k}_{\perp i})$ are then found in terms of relative coordinates ${(x_i,\bm{k}_{\perp i})}$ of the combined system, as defined in~\cref{eq:relcoord}, where
    \begin{equation}
        P^+ = \sum_l  P^{(l)+}
        \,, \quad
        \bm{P}_\perp = \sum_l \bm{P}^{(l)}_\perp
        \, .
    \end{equation}
    This time one can write
    \begin{equation}
        \sket{{\Psi}^{(l)} ({x}_i, {\bm{k}}_{\perp i})} = {\OOPA}{}^{(l)\dagger} \vac \,,
    \end{equation}
    where
    \begin{equation}
    {\OOPA}{}^{(l)\dagger} = \mathcal{C} \,\mathcal{A}^{(l)\dagger} \mathcal{C}^\dagger =
    \mathcal{C}\, \mathcal{B}^{(l)} \widebar{\OOPA}{}^{(l)\dagger}\mathcal{B}^{(l)\dagger} \mathcal{C}^\dagger \,,
    \end{equation}
    with $\mathcal{C}$ implementing the transformation~\eqref{eq:relcoord}.

    \item The wavefunctions $\Psi^{(l)}(x_i,\bm{k}_{\perp i})$ of individual hadrons are combined into a composite state as in~\cref{eq:manybint}:
    \begin{equation}
        \label{eq:lfcombwf}
        {\OOPA}{}^{(1)\dagger} {\OOPA}{}^{(2)\dagger} \ldots \vac \,.
    \end{equation}
\end{enumerate}

Note that the form of transformation~\eqref{eq:boost1} does not imply the factorization of wavefunction~\eqref{eq:lfcombwf}, as operators ${\OOPA}{}^{(l)\dagger}$ do not act on disjoint sets of momenta.
Nevertheless, approximate factorization may still take place if the overlap between the LF wavefunctions $\Psi^{(l)}(x_i,\bm{k}_{\perp i})$ is small.
For this to occur, utilizing a different single-particle basis may be helpful (e.g., using the spatial discretization).

\section{Charges and quantum numbers in the free and interacting theories\label{app:charges}}

Oftentimes quantum numbers arising from the solution of a single-particle Sch\"odinger equation can be naturally associated with known symmetry operators, such as momentum or angular momentum.
In other instances these symmetries are less trivial to identify~-- as in the case of the Laplace–Runge–Lenz vector in the Kepler problem~\cite{shankar2012principles}, or the $Sp(2N,\mathbb{R})$ symmetry of the $N$-dimensional harmonic oscillator~\cite{pan2001exact}.
Formally, given an assignment of real numbers $\lambda_i$ to eigenstates of the Hamiltonian, we can always define the self-adjoint operator $\Lambda$ which multiplies every state by its corresponding $\lambda_i$. Since this operator is by construction diagonal in the eigenbasis of the Hamiltonian, it commutes with the Hamiltonian and is hence the generator of a symmetry~\cite{692404}.
In practice, however, the construction of such operators is not necessary, as one can always interpret quantum numbers of unknown symmetry operators as additional labels of states with fixed eigenvalues of known symmetry operators (the way we usually treat the radial quantum number in spherically-symmetric potentials).

While in the text we assumed that the CSCO of the free and interacting theories share the same charges, one can also consider a more general scenario, in which those differ:
\begin{alignat}{98}
\text{Free CSCO}&\!:\quad&&\{\Hopfree,\,
    &&\chargeop[1],\,
    &&\chargeop[2],\,&&\ldots\} &&,\,\\
\text{Full CSCO}&\!:\quad&&\{\Hopfull,\,
    &&\widetilde{\chargeop}^{(1)},\,
    &&\widetilde{\chargeop}^{(2)},\,&&\ldots\} &&.
\end{alignat}

One can then define a set of intermediate single-particle states of the free theory $\{\sket{\intercollective}\}$, which are the eigenstates of~${\widetilde{\chargeop}}^{(j)}$:
\begin{equation}
    \sket{\intercollective} =
    R_{\intercollective}{}^{\collective}
    \sket{\collective}:
    \quad
    {\widetilde{\chargeop}}^{(j)} \sket{\intercollective}
    =
    \intercollective^{(j)}
    \sket{\intercollective} \, .
\end{equation}

In order to prepare the multi-particle states in the full theory, carrying particular ${\widetilde{\chargeop}}^{(j)}$ charges, one needs to:
\begin{enumerate}
    \item Define operators $\OPad_{\intercollective}$ creating $\{\sket{\intercollective}\}$.
    \item Define Fock states $\sket{\interFF}$ based on the intermediate single-particle states, as well as the operators $\OOPad_{\interFF}$ creating those from vacuum.
    \item Define creation operators in the full theory as
    \begin{equation}
    \label{eq:interAdef}
    \OOPAd_{\interFF}
    \equiv \Wop(\Modeset, \occmax,\maxmodes) \OOPad_{\interFF}\, \Wopd(\Modeset, \occmax,\maxmodes) \, ,
    \end{equation}
\end{enumerate}
where it is crucial that $\Wopd(\Modeset, \occmax,\maxmodes)$ commute with $\chargeop[j]$.
As examples of the described scenario one may consider solving QCD~\cite{Li:2015zda,Jia:2018ary,Mondal:2019jdg} or QED~\cite{Zhao:2014xaa,Vary:2014tqa,Hu:2020arv} in the BLFQ basis~\cite{Vary:2013kma,varybasis}.
In these cases, one is generally interested in the states of definite momentum in the interacting theory.
However, for the sake of improving numerical convergence, one may choose to define Fock states using as single-particle states the eigenbasis of a confining potential.

\section{DLCQ Hamiltonian of the $\phi^4_{1+1}$ theory\label{app:phi4ham}}

The $\phi^4_{1+1}$ theory was first studied within the framework of Discretized Light-Cone Quantization (DLCQ) in Ref.~\cite{phi4in2d}.
The Lagrangian of the theory is given by
\begin{equation}
    \mathcal{L}
    =  \frac{1}{2}\partial^{\mu}\phi\partial_\mu\phi- \frac{1}{2}\mass^2\phi^2-\frac{\lambda}{4!}\phi^4 \,,
\end{equation}
where the metric is chosen as~$g=\diag\{1,-1\}$.
By means of the coordinate transformation
\begin{equation}
    \label{coords}
    x^\pm = x^0 \pm x^1
    \, ,
\end{equation}
one switches to the so-called light-front time ($x^+$) and distance ($x^-$).
The corresponding metric is ${g^{++}=g^{--}=0}$, ${g^{+-}=g^{-+}=2}$.

The evolution in the~$x^+$ direction is governed by the Hamiltonian
\begin{equation}
\label{eq:pminus}
\begin{alignedat}{8}
    \mathcal{P}^-
    &= &&\int_{-L}^{+L} &&\d x^- \mathcal{P}^-
    \\&= &&\int_{-L}^{+L} &&\d x^- \left(\dfrac{1}{2}\mass^2\phi^2+\dfrac{\lambda}{4!}\phi^4\right)
    = \dfrac{L}{2\pi} \mathcal{H} \,,
\end{alignedat}
\end{equation}
where $L$ defines the box size, ${x\in[-L,\,L]}$.
However, throughout the paper we refer to $\mathcal{H}$ as the Hamiltonian of the model.

The longitudinal light-cone momentum operator is given by
\begin{equation}
\label{eq:pplus}
    \mathcal{P}^+ = \dfrac{1}{2} \int_{-L}^{+L} \d x^- \partial^+\phi\,\partial^+ \phi
    =\dfrac{2\pi}{L}\mathcal{K} \, ,
\end{equation}
where $\mathcal{K}$ is the dimensionless longitudinal momentum operator, also known as \emph{harmonic resolution}.

The plane wave expansions of the free field in DLCQ acquires the following form:
\begin{equation}
\label{eq:philf}
\begin{gathered}
    \phi(x^\mu) = \sum\limits_{\Moden=1}^\infty \dfrac{1}{\sqrt{4\pi \Moden}}
    \Bigl( \OPa_\Moden \me^{-\iu \Modep^\mu_\Moden x_\mu} + \OPad_\Moden \me^{\iu \Modep^\mu_\Moden x_\mu} \Bigr) \, , \\
    \Modep^+_\Moden = \dfrac{2\pi}{L} \Moden
    \, ,\quad
    \Modep^-_{\Moden} = \dfrac{\mass^2}{\Modep^+_\Moden}
    \, ,\quad
    \Moden=1,2,3\ldots
\end{gathered}
\raisetag{5\baselineskip}
\end{equation}
Note that in~\cref{eq:philf} we assumed the periodic boundary conditions and neglected the contribution of zero modes~\cite{BRODSKY1998299,Heinzl:1995xj,Harindranath:1999vf,Vary:2021cbh}.

Upon substituting the free field expansion~\eqref{eq:philf} into~\cref{eq:pminus} and~\cref{eq:pplus}, the normal-ordered expression for the harmonic resolution operator and the Hamiltonian acquire the following form:
\begin{equation}
    \mathcal{K} = \sum \limits_\Moden \Moden \, \OPad_\Moden \OPa_\Moden \, ,
\end{equation}
and
\begin{equation}
\label{eq:phi4hamapp}
    \mathcal{H}_{\text{full}} = \mathcal{H}_{\text{free}} + \mathcal{H}^{I} \, ,
\end{equation}
where
\begin{equation}
    \mathcal{H}_{\text{free}} = \sum_{\Moden}\frac{1}{\Moden}\OPad_\Moden \OPa_\Moden\left(\mass_B^2+\dfrac{\lambda}{4\pi}\frac{1}{2}\sum_{\Modek=1}^{\infty}\frac{1}{\Modek}\right) \,,
\end{equation}
and
\begin{equation}
\label{eq:HI}
      \begin{alignedat}{9}
\mathcal{H}^{I}&=\frac{1}{4}\frac{\lambda}{4\pi}\sum_{\Modek \Model  \Modem\Moden}\frac{\OPad_\Modek \OPad_\Model \OPa_\Modem \OPa_\Moden}{\sqrt{\Modek\Model\Modem\Moden}}\delta_{\Modem+\Moden,\Modek+\Model}
     \, ,\\
&+\frac{1}{6}\frac{\lambda}{4\pi}\sum_{\Modek\Model\Modem\Moden}\frac{\OPad_\Modek \OPa_\Model \OPa_\Modem \OPa_\Moden}{\sqrt{\Modek\Model\Modem\Moden}}\delta_{\Modek,\Modem+\Moden+\Model}\\
&+\frac{1}{6}\frac{\lambda}{4\pi}\sum_{\Modek\Model\Modem\Moden}\frac{\OPad_\Moden \OPad_\Modem \OPad_\Model a_\Modek}{\sqrt{\Modek\Model\Modem\Moden}}\delta_{\Modek,\Modem+\Moden+\Model} \, .
    \end{alignedat}
\end{equation}

The divergent sum in the free Hamiltonian is known as ``self-induced intertia''. It arises due to normal ordering and is ignored in the main text.

\section{Interaction turn on via unitary coupled cluster\label{app:ucc}}
In this Appendix, we describe an algorithm for constructing of the adiabatic turn-on operator, using  the unitary coupled cluster (UCC) method~\cite{shen2017quantum,Romero_2018,filip2020stochastic}.
As an in~\Cref{sec:dlcq}, we use $\phi^4_{1+1}$ theory for illustrating our ideas, yet keep the arguments general.

In order to define $\Wop[K]$ as in~\cref{eq:act,eq:unit,eq:vset,eq:actprime}, we write it in the form ${\Wop[K] = \me^{-\iu \Vop[K]}}$, where the Hermitian operator $\Vop[K]$ is sought as a polynomial in creation and annihilation operators.
The coefficients in this polynomial are variables, which will be found via solving a linear system of equations.
Na\"ively, one would expect these equations to stem from the condition that, in the basis of $\bigl\{\Fock_K\bigr\}$, the matrix elements of $\Wopd[K]$ coincide with those of $\W[K]$, the matrix of eigenvectors of the full Hamiltonian relating the bases of the free and interacting theory.
While conceptually correct, this approach requires two adjustments.
First of all, in order to linearize the problem of finding the coefficients in $\Vop[K]$, instead of equating the matrix of $\Wop[K]$ to $\W[K]$, we are going to directly equate the matrix of $\Vop[K]$ to $\VV[K] = \iu \ln (\W[K])$.
Second, in order for $\Wop[K]$ to act as adiabatic turn-on in all the sectors of $K^\prime<K$ as well, we additionally require that the matrix of $\Vop[K]$ in the basis of any $\{\Fock_{K^\prime<K}\}$ matches with $\VV[K^\prime]$.

We begin by splitting $\Vop[K]$ into the ``diagonal'' and ``non-diagonal''parts:
\begin{equation}
    \label{eq:VK}
    \Vop[K] = \Vop[K][\mathrm{d}] + \Vop[K][\rm n/d] \, .
\end{equation}
The ``diagonal'' part has the following form:
\begin{equation}
\label{eq:VKdsum}
\Vop[K][\rm d] = \sum_{r = 1}^K \Vop[K,\,r][\mathrm{d}]
\end{equation}
Operators $\Vop[K,\,r][\mathrm{d}]$ contain monomials assembled from creation and annihilation operators acting on the same set of $r$ particles:
\begin{equation}
\begin{alignedat}{9}
    \label{eq:VKdterms}
    \Vop[K,\,1][\mathrm{d}] &=
    \theta^{\mathrm{d}}_{1,1} \OPad_1\OPa_1 +
    \ldots &&+
    \theta^{\mathrm{d}}_{1,{{K}\choose{1}}} \OPad_K\OPa_K  \, ,\\
    \Vop[K,\,2][\mathrm{d}] &=
    \theta^{\mathrm{d}}_{2,1} \OPad_1\OPad_1\OPa_1\OPa_1 &&+
    \theta^{\mathrm{d}}_{2,2} \OPad_1\OPad_2\OPa_1\OPa_2
    \\ &
    &&\mathllap{+\ldots}+ \theta^{\mathrm{d}}_{1,{{{K}\choose{2}}}} \OPad_K\OPad_K\OPa_K\OPa_K \ \\
    \ldots
    \\
    \Vop[K,\,K][\mathrm{d}] &=
    \mathrlap{
    \theta^{\mathrm{d}}_{K,1} \OPad_1\ldots\OPad_1\OPa_1\ldots\OPa_1
    }
    \\&
    \hspace{1.5cm}+
    \mathrlap{
    \theta^{\mathrm{d}}_{K,{{K}\choose{K}}} \OPad_K\ldots\OPad_K\OPa_K\ldots\OPa_K \, .
    }
\end{alignedat}
\end{equation}
Note that
\begin{equation}
    \label{eq:product_def}
    \Vop[K,\,r][\mathrm{d}] =
    \bigl[
    \Vop[K,\,r-1][\mathrm{d}] \times
    \Vop[K,\,1][\mathrm{d}]
    \bigr] \, ,
\end{equation}
where by $[\ldots\times\ldots]$ we denote the sum of all terms arising in the normal-ordered product, with coefficients being the new independent parameters.
By construction, $\Vop[K][\mathrm{d}]$ contains terms which do not change the particle content of Fock states.

In order to define $\Vop[K][\rm n/d]$, it is convenient to first introduce a \emph{minimal} operator $\oprelat{K,i}{K,j}$ connecting the two Fock states $\snfree[K][j]$ and $\snfree[K][i]$:
\begin{equation}
    \label{eq:connect}
    \oprelat{K,i}{K,j} \snfree[K][j] = \snfree[K][i] \, ,
\end{equation}
where it is assumed that $\oprelat{K,i}{K,j}$ is a monomial consisting of the smallest number of single-mode creation and annihilation operator.
$\oprelat{K,i}{K,j}$ is a normal-ordered product of operators annihilating particles present in $\snfree[K][j]$ but not in $\snfree[K][i]$ and operators creating particles present in $\snfree[K][i]$ but not in $\snfree[K][j]$.
The basis states in~\cref{eq:connect} are normalized to $1$ while the coefficient ensuring the correct normalization is stored in the definition of the operator.

Next, we then write $\Vop[K,\,r][\mathrm{n/d}]$ as a following sum:
\begin{equation}
\label{eq:VKndsum}
\Vop[K][\mathrm{n/d}] = \sum_{r = 2}^K \Vop[K,\,r][\mathrm{n/d}] \, ,
\end{equation}
where
\begin{equation}
\label{eq:VKndterms}
\begin{alignedat}{9}
    \Vop[K,\,2][\mathrm{n/d},\, 2] &= \sum_{i\neq j} \theta^{\mathrm{n/d}}_{K,i,j}\oprelat{K,i}{K,j} +
    (\theta^{\mathrm{n/d}}_{K,i,j})\oprelatd{K,i}{K,j}
    \, , \\
    \Vop[K,\,r][\mathrm{n/d}]
    &= \bigl[ \Vop[K,\,\mathrm{n/d},\, r-1] \times \Vop[K,\,\mathrm{d},\, 1] \bigr] \, .
\end{alignedat}
\end{equation}
The term $\Vop[K,\,1][\mathrm{n/d}]$ is absent in the sum~\eqref{eq:VKndsum}, because in $\phi^4$ theory any two states belonging to the block of the same $K$ have to differ by at least two particles.

In order for $\Vop[K]$ to be Hermitian, the coefficients in~\cref{eq:VKdterms} have to be real, while those in~\cref{eq:VKndterms} could be, generally, complex.
In fact, for the considered model, $\Vop[K][\mathrm{d}]$ vanishes and the coefficients in $\Vop[K][\mathrm{n/d}]$ are purely imaginary, which explains why a common definition of UCC, $\me^{T-T^\dagger}$, does not include the imaginary unit in the exponent~\cite{McClean_2016}.

\bigskip

We now describe the algorithm for expanding an arbitrary operator $\Vop$ in terms of creation and annihilation operators, so that its matrix elements in the basis of $\{\Fock\}$ would be given by matrix $\VV[]$.

I. Diagonal terms.
\begin{enumerate}
    \item Set  $r_{\mathrm{max}} = 1$.
    \item Truncate the sum in eq.~\eqref{eq:VKdsum} as
    \begin{equation}
    \Vop[K][\rm d] = \sum_{r = 1}^{r_{\mathrm{max}}} \Vop[K,\,r][\mathrm{d}] = \sum_j^{J^{\mathrm{d}}_r} \theta^{\mathrm{d}}_j v^{\mathrm{d}}_j \, ,
    \end{equation}
    where $v_j$ stand for individual monomials in creation and annihilation operators whose total number is~$J^{\mathrm{d}}_r$.

    \item Form a matrix $A^{\mathrm{d}}_{ij}$ and a vector $h_j$:
    \begin{subequations}
    \begin{gather}
        A^{\mathrm{d}}_{ij} = \sand{\mathcal{F}_i}{v^{\mathrm{d}}_j}{\mathcal{F}_i} \, ,\\
        h^{\mathrm{d}}_i = \VV[{ii}],\\
        \nonumber
        i \in [ 1,\, \bigl|\{\Fock\}\bigr| ] \, ,\quad
        \nonumber
        j \in [ 1,\, J^{\mathrm{d}}_r ] \, .
    \end{gather}
    \end{subequations}
    \item Determine whether solutions to the system
    \begin{equation}
    \label{eq:dsys}
        \sum_{j} A^{\mathrm{d}}_{ij} \theta^{\mathrm{d}}_j = h^{\mathrm{d}}_i
    \end{equation}
    exist.
    According to the Rouch\'e-Capelli theorem~\cite{andreescu2016essential}, the linear system~\eqref{eq:dsys} has solutions iff the rank of the augmented matrix $(A^{\mathrm{d}}_{ij} | h^{\mathrm{d}}_i)$ is no larger than the rank of the augmented matrix $A^{\mathrm{d}}_{ij}$.
    The solution to~\eqref{eq:dsys} with the least Euclidean norm is found from the following minimization problem~\cite{datta2010numerical}:
    \begin{equation}
        \theta^{\mathrm{d}}_j = \min_{x_j} \Bigl\|\sum_{ij}A^{\mathrm{d}}_{ij} x_j - h^{\mathrm{d}}_i\Bigr\| \, .
    \end{equation}

    If $\rank A^{\mathrm{d}}_{ij} < \rank (A^{\mathrm{d}}_{ij} | h^{\mathrm{d}}_i) $, increase $r_{\mathrm{max}}$ by one and return to step 2.
\end{enumerate}

II. Non-diagonal terms.
\begin{enumerate}
    \item Set  $r_{\mathrm{max}} = 2$.
    \item Truncate the sum in~\cref{eq:VKndsum} as
    \begin{equation}
    \Vop[K][\rm n/d] = \sum_{r = 1}^{r_{\mathrm{max}}} \Vop[K,\,r][\mathrm{n/d}] = \sum_j^{J^{\mathrm{n/d}}_r} \theta^{\mathrm{n/d}}_j v^{\mathrm{n/d}}_j \, ,
    \end{equation}
    where $v^{\mathrm{n/d}}_j$ stand for individual monomials in creation and annihilation operators.
    \item Form a matrix $A^{\mathrm{n/d}}_{ij}$ and a vector $h^{\mathrm{n/d}}_j$:
    \begin{subequations}
    \begin{gather}
        A^{\mathrm{n/d}}_{ij} = \sand{\mathcal{F}_{i_1}}{v^{\mathrm{n/d}}_j}{\mathcal{F}_{i_2}} \, , \\
        h^{\mathrm{n/d}}_i = V_{i_1i_2} \, , \\
        \nonumber
        i = (i_1, i_2) \, , \quad i_1 > i_2 \, ,\\
        \nonumber
        i_1, i_2 \in \bigl[1, \,\bigl|\{\Fock\}\bigr| \bigr]\\
        \nonumber
        i = \bigl[1,\,\bigl|\{\Fock\}\bigr|\bigl(\bigl|\{\Fock\}\bigr|-1\bigr)/2\bigr]\ ,\\
        \nonumber
        j = [1,\,J^{\mathrm{n/d}}_r] \, .
    \end{gather}
    \end{subequations}
    \item Determine whether the linear system
    \begin{equation}
        \sum_{j} A^{\mathrm{n/d}}_{ij} \theta^{\mathrm{n/d}}_j = h^{\mathrm{n/d}}_i
    \end{equation}
    has solutions. If $\rank A^{\mathrm{n/d}}_{ij} < \rank (A^{\mathrm{n/d}}_{ij} | h^{\mathrm{n/d}}_i) $, increase $r_{\mathrm{max}}$ by one and return to step 2. Otherwise find the solutions $\theta^{\mathrm{n/d}}_j$ via solving the linear system:
    \begin{equation}
        \theta^{\mathrm{n/d}}_j = \min_{x_j} \Bigl\|\sum_{ij}A^{\mathrm{n/d}}_{ij} x_j - h^{\mathrm{n/d}}_i\Bigr\| \, .
    \end{equation}
\end{enumerate}
By construction, the algorithm above produces an operator polynomial in creation and annihilation operators of lowest possible degree, such that its matrix elements in the basis of Fock states are given by the provided matrix.
In application to our problem, we choose the matrix $\VV[]$ to be a block-diagonal matrix comprising blocks of harmonic resolution up to $K$:
\begin{equation}
        \VV[] = \diag \{
        \VV[2,\mathrm{odd}],
        \VV[2,\mathrm{even}],
        \ldots,
        \VV[K,\mathrm{odd}],
        \VV[K,\mathrm{even}]
        \} \, ,
\end{equation}
acting on states
\begin{equation}
        \begin{alignedat}{8}
        \{\Fock\} =
        &\{ \Fock[2,\mathrm{odd}] \} \cup
        \{ \Fock[2,\mathrm{even}] \}\\
         \cup &\ldots \cup
        \{ \Fock[K,\mathrm{odd}] \} \cup
        \{ \Fock[K,\mathrm{even}] \}
        \, .
        \end{alignedat}
\end{equation}

\section{Quantum Implementation of the Creation Operator\label{app:creat}}

Below we review several ways of implementing the action of the creation operator in a quantum circuit.
We consider the calculation of a matrix element
\begin{equation}
\label{eq:aaa}
    \sand{f}{ \ldots \OPad_{\collective_2} U_2 \OPad_{\collective_1} U_1}{i} \,.
\end{equation}

\begin{enumerate}
    \item If the direct encoding is used, and each $\OPad_{\collective_j}$ is mapped onto polynomially many Pauli terms~\cite{bravyi2002fermionic,seeley2012bravyi,sawaya2020resource}, those can be readily substituted into~\cref{eq:aaa}, resulting in circuits analogous to that one in~\Cref{fig:K1K2}, with creation operators replaced with Pauli operators.
    The number of measurements then grows exponentially in the number of creation operators in~\cref{eq:aaa}.

    \item Under the same conditions on $\OPad_{\collective_j}$ as in 1. above, one can start off the explicit form of the operator matrix, and implement its action on a quantum state by constructing its block encoding using such algorithms as linear combination of unitaries~\cite{gui2006general,childs2012hamiltonian} or FABLE~\cite{camps2022fable}.

    \item For fermions or bosons with $\occmax=1$ one can implement the non-unitary action of the creation and annihilation operators by performing a measurement on the qubit storing the wavefunction of the quantum state.
    For example, in order to implement the operator of the form ${X_j-\iu Y_j}$, one would measure the $j$th qubit, and disregard the state if the qubit is found in the $0$ state or flip the qubit using the $X$ gate and continue the computation.
    In order to account for the renormalization of amplitudes in the quantum state due to the measurement, one would need to multiply the final result by the fraction of successful measurements of the $j$th qubit.
    A more general version of this approach is described in~\cite{PhysRevA.101.012330}.

    \item Implementing the action of creation and annihilation operators in the compact encoding scheme has been discussed in Refs~\cite{Kreshchuk:2020dla,kirby2021quantum}.
\end{enumerate}
\end{appendices}

\end{document}